\documentclass[aps, prd, showpacs, twocolumn, superscriptaddress, groupedaddress, floatfix]{revtex4-2}

\usepackage{graphicx}
\usepackage{yfonts}
\usepackage{float}
\usepackage{amsmath}
\usepackage{amssymb}
\usepackage{amsfonts}
\usepackage{amsthm}
\usepackage{mathtools}
\usepackage{dcolumn}
\usepackage{listings}
\usepackage{subfigure, rotating, bm, array}
\usepackage[utf8]{inputenc}
\usepackage[english]{babel}
\usepackage[pagebackref=false, colorlinks=true]{hyperref}
\hypersetup{
linkcolor=blue,     
citecolor=blue,     
urlcolor=blue}      

\usepackage{float}

\begin{document}
\title{Gravitational collapse of matter in the presence of Quintessence and Phantom-like scalar fields}
\author{Priyanka Saha}
\email{priyankas21@iitk.ac.in}
\affiliation{Department of Physics, Indian Institute of Technology, Kanpur,
Kanpur-208016, India}
\author{Dipanjan Dey}
\email{deydipanjan7@gmail.com}
\affiliation{Department of Mathematics and Statistics, Dalhousie University, Halifax, Nova Scotia, Canada B3H 3J5.}
\author{Kaushik Bhattacharya}
\email{kaushikb@iitk.ac.in}
\affiliation{Department of Physics, Indian Institute of Technology, Kanpur,
Kanpur-208016,India}

\begin{abstract}
In this work, we propose a model of the gravitational collapse of dark matter in the presence of quintessence or phantom-like scalar fields. Our treatment is based on the principles of general relativity up to virialization. We have chosen a spherical patch that starts to collapse gravitationally as it happens in top-hat collapse. It is seen that although the dark matter sector collapses the dark energy sector does keep a profile that is almost similar to the dark energy profile for the background expanding Friedmann-Lemaitre-Robertson-Walker (FLRW) universe for suitable model parameters. It is observed that in order to formulate the problem in the general relativistic setting one has to abandon the idea of a closed FLRW isolated collapsing patch. General relativity requires an external generalized Vaidya spacetime to be matched with the internal spherical patch whose dynamics is guided by the FLRW metric. It is shown that almost all collapses are accompanied by some flux of matter and radiation in the generalized Vaidya spacetime. Some of the spherical regions of the universe are seen not to collapse but expand eternally, producing void-like structures. Whether a spherical region will collapse or expand depends upon the initial values of the system and other model parameters. As this work shows that collapsing structures must emit some form of radiation, this may be taken as an observational signature of our proposal.
\end{abstract}
\maketitle

\section{Introduction}
 The formation of structure in a homogeneous and isotropic universe is always an interesting and evergreen topic in astrophysics and cosmology. In the standard picture, the seed for structure formation in cosmology comes from linear perturbation theory. During or after recombination the cosmological perturbations,  for some modes, start to grow and does not remain strictly linear. These modes act as seeds for future structure formation in the universe. Some of these perturbation modes move out of the linear paradigm and enter the nonlinear mode where different physical principles are operational. Just before entering the nonlinear regime Jeans instability \cite{Jeans:1902fpv,Moradpour:2019wpj} and other effects guide the formation of structures. Gravitationally bound structures, from the cluster of galaxies scales to much lower scales, are supposed to have been born due to nonlinear instabilities.  In the standard picture of structure formation, it is assumed that primarily the dark matter sector plays the most important role. The dark matter sector is supposed to be composed of a fluid with zero pressure which follows the gravitational potential produced by a marginally denser region and tries to collapse about those regions. The baryonic matter follows the dark matter flow \cite{Paddy, Primack, Dey1, Dey3}. One of the most important semi-relativistic methods used to study structure formation is called the top-hat collapse \cite{Gunn}. In this collapse process, it is assumed that if in some closed region of the cosmos, the density of dark matter has exceeded the background matter density then a collapse follows. In top-hat collapse, the closed overdense region at first expands following the background expansion, but this expansion halts at a certain moment due to gravity and there is a turnaround. Following the turnaround, the closed region starts to collapse. A pure general relativistic top-hat collapse generally produces a singularity as the end state, since the collapsing matter is homogeneous and dust-like \cite{Oppenheimer:1939ue}. However, in astrophysics, it is assumed that much before the formation of a singularity the collapsing fluid virializes. The virialized end state of the collapse signifies structure formation. In this sense, the top-hat collapse is a semi-relativistic process where people use a semi-Newtonian paradigm to interpret the end phase of the collapse.

Traditionally one does not take into account the role of the cosmological constant, $\Lambda$, in the structure formation process. Some authors have tried to incorporate the effects of such a constant in the gravitational collapse process \cite{Garfinkle:1990jr,Deshingkar:2000hd,Sharif:2006bp,Markovic:1999di}.  Traditional $\Lambda$CDM models have their own difficulties \cite{DelPopolo:2016emo,Perivolaropoulos:2021jda}, and consequently, the dynamical dark energy models based on scalar fields have been introduced. One of the most widely used scalar fields in this paradigm is the quintessence field. Phantom-like scalar fields, with a negative kinetic term, also is used to model dark energy \cite{Bamba:2012cp,Caldwell:1999ew,Carroll:2003st,Copeland:2006wr,Caldwell:2003vq}. In this paper, we will mainly be working with these two types of scalar fields.  Our main goal is to study the gravitational collapse process in a two-component universe, with dark matter and a scalar field acting as source of dynamic dark energy. Many authors have attempted such a problem in various forms \cite{Herrera:2019nwd,Chang:2017vhs,Mota:2004pa,Steinh,Weinberg:2002rd}. In almost all of the attempts the authors never used a formal general relativistic approach although they used one or two equations that can only be found in a general relativistic setting. The main reason for such a purely phenomenological approach by the previous authors is primarily based on the following reason. If one wants to apply general relativistic treatment for the gravitational collapse of a closed spherical region then one has to start with the clsoed Friedmann-Lemaitre-Robertson-Walker (FLRW) spacetime with some matter inside the closed region. This closed region does not exchange energy with the outside as this spacetime is assumed to be 'closed' and acts as an isolated system. If this closed region undergoes a gravitational collapse, as in top-hat collapse, then the energy density of the matter inside grows as the region shrinks as that is the only way the energy of matter can be conserved.  On the other hand in a two-component system, where one of the components can be a dark energy candidate, this logic may not apply as the dark energy sector may remain homogeneous and unclustered. By unclustered dark energy, we mean that the energy density of dark energy practically remains the same as that of the expanding background FLRW spacetime. In simple terms, the dark energy sector may not collapse at all following the dark matter partner inside the closed region. In such a case energy conservation becomes problematic and the problem becomes paradoxical. To evade this problem, previous authors have used a pure phenomenological method.  In this method, one does not perceive the problem relativistically, where one uses an FLRW metric with a positive spatial curvature constant and then writes down the Friedmann equations. The first Friedmann equation (containing the square of the first derivative of the local scale factor) particularly becomes problematic as it requires an estimate of all the known energy sources inside the spherical patch. As energy may not be conserved, this equation becomes redundant. Mostly all of the previous works in this field only use the other Friedmann equation containing the second derivative of the scale factor and consider it as a second-order ordinary differential equation in time and solve it with appropriate initial conditions.   

We have addressed the above-mentioned problem in a more relativistic way. As it is known that for unclustered dark energy, the scalar field sector does not collapse, we expect that this sector primarily leaks out of the boundary of the closed, positively curved spacial region. To incorporate such an idea we match the internal FLRW patch with a generalized Vaidya spacetime before the internal spacetime closes (the internal radial distance marker is less than one). As a result of this we predict the emission of radiation from the boundary of the collapsing region, this radiation is naturally obtained in generalized Vaidya spacetimes. The two spacetimes are matched at a time-like hypersurface using the standard junction conditions of general relativity. The matching of the spacetimes solves the issue of non-conservation of energy in the closed patch as in the modified scenario the spherical patch is radiating energy outside and ideally does not remain an isolated patch anymore. In our model the collapsing dark matter cloud affects the dark energy density locally as the spherical region under collapse forces the dark energy sector to radiate. This model is natural in the sense that the effect of a gravitational collapse does not go unnoticed in the dark energy sector, it reacts to the collapse by transforming locally into radiation although its energy density follows the energy density of the background spacetime. We think this is the first serious attempt to produce a formal general relativistic paradigm of the spherical collapse of dark matter in the presence of dark energy.

Although we have tried to formally establish a general relativistic attempt to tackle the problem of spherical collapse of matter in presence of dark energy we do not fully extend the relativistic formalism up to the formation of the singularity which is inevitable in such situations. The primary reason for using a more phenomenological process, to end the collapse,  is related to the fact that large scale structures exist and perhaps they are produced from some virialization process. Our radiating collapsing structure inevitably must virialize at some point of time and after that time the system does not remain relativistic. Virialization by itself is not built in the collapsing process, one has to bring in this pseudo-Newtonian concept to explain the existence of large-scale gravitationally bound objects. In the cases of collapse in the presence of unclustered dark energy, the dark matter sector primarily collapses and virializes whereas the dark energy sector does not virialize. Although the dark energy sector does not virialize it does affect the virialization process of the dark matter sector. 

In all the cases of spherical collapse, which we have studied in this paper, the dark energy component remains primarily unclustered and homogeneous for some suitable small values of parameters in our model. Our attempt to study collapse in such two-component systems does not only produce unclustered dark energy, in some situations, it is seen that the closed, spherical region does not proceed to a collapse at all. In these cases, we have an eternal expansion of a small local spherical patch in the background of the spatially flat FLRW spacetime. These regions act like voids as the matter density inside them decreases. The dark energy density in these patches exceeds the dark energy density of the background and consequently, we can say that clustered dark energy can also be produced in our model. Whether a spherical patch will end up in a virialized state or an expanding phase depends upon the parameters of the theory and initial conditions.  In these expanding regions the dark matter density remains a factor of ten smaller than the background spacetime for some time. Ultimatly as these patches expand the matter density drops. The dark energy density remains less than the background dark energy density for quintessence fields. For phantom fields the dark energy desnity in the spherical patch tends to be more in the voids.  

The work in this paper is organized in the following way. In section \ref{sec2} we elaborately discuss the semi-Newtonian theory of virialization in a two component universe where the two components are related to the dark matter sector and the dark energy sector. We call this treatment semi-Newtonian as we use the language of Newtonian potentials although the energy conservation equations are obtained from an expanding universe paradigm. In section \ref{sec3} we present the general relativistic formalism for our work. This section contains the junction conditions used to join a collapsing/expanding closed FLRW spacetime with the generalized Vaidya spacetime. Section \ref{sec4} presents the basic equations which guides the collapse of a spherical FLRW patch in presence of a quintessence/phantom like scalar field and dark matter. In section \ref{sec5} we have presented the results obtained from the calculations in the previous section. This section shows the details of the various collapsing processes. Section \ref{sec6} gives a summary of the work presented in this paper.  
\section{Virialization state of dark matter in the presence of dark energy}
\label{sec2}
The total gravitational potential of the over-dense region  of a two-fluid system consisting of dark matter (DM) and dark energy (DE) can be written as \cite{Maor:2005hq}
\begin{eqnarray}
V_T = \frac12 \int_v &\rho_{DM}&\phi_{DM} ~dv + \frac12\int_v \rho_{DM}~\phi_{DE} ~dv \nonumber\\ &+& \frac12\int_v \rho_{DE}~\phi_{DM} ~dv+ \frac12\int_v \rho_{DE}~\phi_{DE}~ dv\,\, ,\nonumber\\
\label{VT1}
\end{eqnarray}
where $\phi_{DM}$ and $\phi_{DE}$ are the gravitational potentials of dark matter and dark energy, respectively, and $\rho_{DM}$ and $\rho_{DE}$ are the energy density of dark matter and dark energy, respectively. The integration is done over the whole volume ($v$) of the spherical over-dense region. The non-zero values of the four integrations written above can be used to classify the two-fluid system into the following four distinguishable scenarios,
\begin{itemize}
\item In the first scenario, the dark energy effect is totally neglected considering only the first integration in Eq.~(\ref{VT1}) is non-zero. In this case, the spherical over-densities of dark matter behave like an isolated sub-universe and virialize at a certain radius \cite{Gunn}. 
\item If only the first two integrations in Eq.~(\ref{VT1}) contribute to the total gravitational potential of the over-dense region, then it can be shown that there exists a non-negligible effect of dark energy which affects the virialization process of the spherically symmetric over-dense regions of dark matter. In this scenario, dark energy cannot cluster and virialize with dark matter, and therefore, the dark energy density inside the over-dense region is similar to the external dark energy density. Hence, this type of model is known as the homogeneous dark energy model \cite{Lahav:1991wc, Steinh, Shapiro, Horellou:2005qc}.
\item In the third scenario, dark energy does not virialize with dark matter though it can cluster inside the over-dense regions. In this scenario, it is considered that from the starting point of the matter-dominated era, dark energy moves synchronously with the dark matter on both the Hubble scale and the galaxy cluster scale. This scenario is known as the clustered dark energy scenario \cite{Basilakos:2003bi, Maor:2005hq, Basilakos:2006us, Basilakos:2009mz, Chang:2017vhs, Dey2}.
\item At last, in the fourth scenario, dark energy can cluster and also virializes with dark matter inside the spherical over-dense regions \cite{Maor:2005hq}.
\end{itemize}
\begin{figure}[h!]
\centering
{\includegraphics[width=90mm]{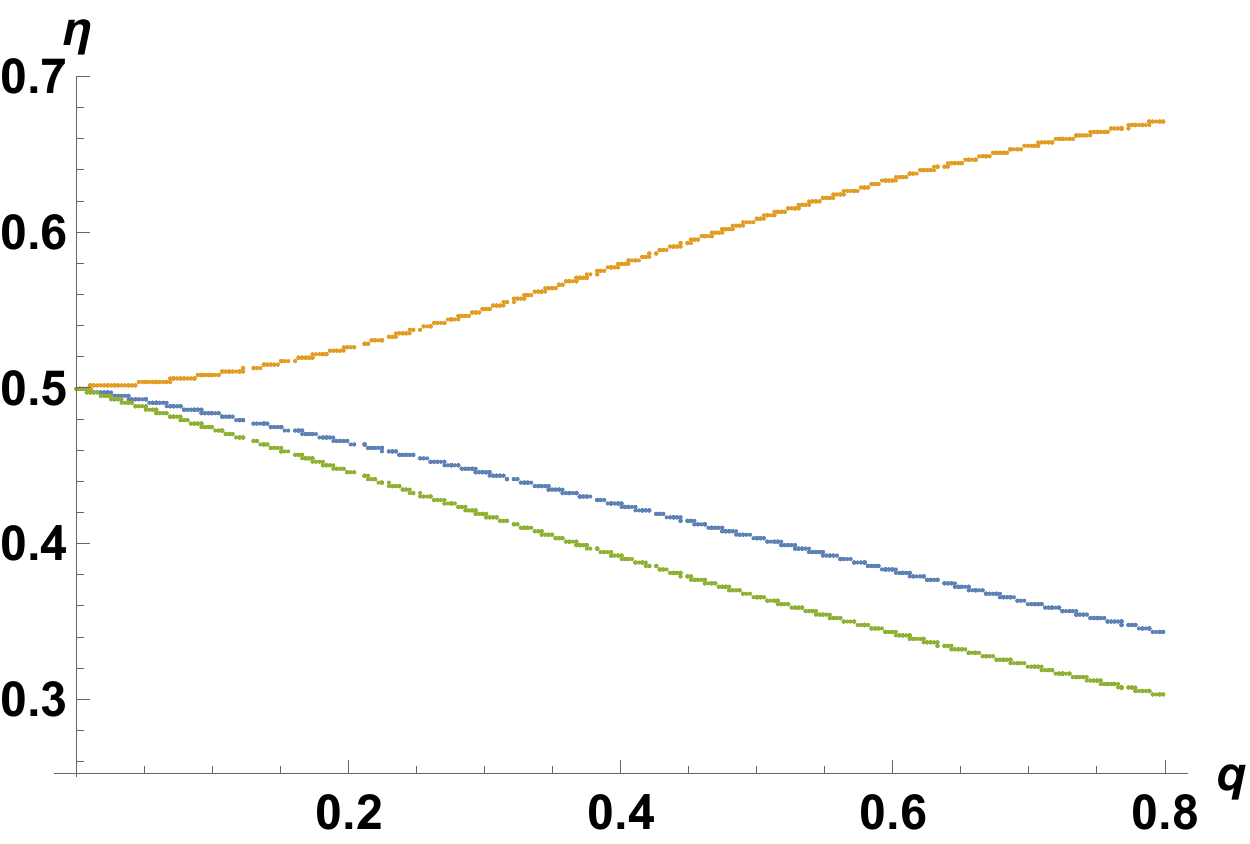}}
\caption{Figure shows how $\eta$ varies with $q$ (where $\eta=\frac{R_{vir}}{R_{max}}$  $q=\left(\frac{\rho_{DE}}{\rho_{DM}}\right)_{t=t_{max}}$) for following three scenarios: a) dark energy can cluster and virialize with dark matter (depicted by brown curve) b) dark energy can cluster and cannot virialize inside the over dense region (depicted by green curve) c) homogeneous $\Lambda$ dark energy (depicted by blue curve).}
\label{etavsq1}
\end{figure}
If we consider no influence of dark energy in the evolution of the dark matter over-densities, then as mentioned previously, the first integration of Eq.~(\ref{VT1}) contributes to the total gravitational potential of the over-dense regions. This scenario is described by the top-hat collapse model, where one self-gravitating fluid, inside a spherical over-dense region, virializes \cite{Gunn}. In the top-hat collapse model, the over-dense region expands first with the background but at a slower rate than that of the background, and then after a certain turnaround radius ($R_{max}$), the over-dense region starts collapsing. At the turnaround radius, momentarily the kinetic energy of the over-dense region becomes zero, and the total gravitational potential energy ($V_T$) of the region becomes the total energy ($E_T$) of the same at that moment. The total energy inside the spherical over-dense region, when it reaches the turnaround radius ($R_{max}$), is:
\begin{eqnarray}
E_T|_{t = t_{max}} = V_T = \frac12\int_{v_{max}} &\rho_{DM}&\phi_{DM} ~dv = -\frac{3M^2}{5R_{max}}\,,\nonumber\\
\end{eqnarray} 
where $t_{max}$ is the turnaround time. 
At the virialization time $t = t_{vir}$, the total kinetic energy of the over-dense region $E_{KE}|_{t_{vir}}= - \frac{V_T|_{t_{vir}}}{2} $. Therefore, at the virialization time, the total energy of the over-dense region $E_{T}|_{t_{vir}} = \frac{V_T|_{t_{vir}}}{2}$. Hence, using energy conservation, one can show that the spherically symmetric over-densities virialize when $\eta= \frac{R_{vir}}{R_{max}}=0.5$. In order to model the dynamics of the over-dense region, if one uses the closed FLRW spacetime, then it can be shown that $t_{vir}=1.81 t_{max}$.

In \cite{Lahav:1991wc} and \cite{Steinh}, the authors investigated the cosmological scenario where the dark energy is homogeneous, i.e., the internal and external dark energy densities are the same. In \cite{Lahav:1991wc}, the authors studied the effect of the cosmological constant on the virialization of the spherical over-densities, whereas in  \cite{Steinh}, the authors consider the homogeneous quintessence  dark energy model. As previously mentioned, in the homogeneous dark energy scenarios, the dark energy does not cluster and virialize inside the spherical over-densities of dark matter, however, the virialization process of the over-densities is modified since there exist a non-zero energy density and negative pressure of dark energy, and this effect of dark energy can be realized by the different values of $\eta$. For the homogeneous dark energy scenario, the total potential energy of the over-dense region can be written as \cite{Maor:2005hq}
\begin{eqnarray}
V_T = \int_v &\rho_{DM}&\phi_{DM} ~dv + \int_v \rho_{DM}~\phi_{DE} ~dv\,\, ,
\end{eqnarray}
which gives
\begin{eqnarray}
V_T=-\frac{3M^2}{5R}\left[1 - ~\frac{q}{2}(1+3\omega)\left(\frac{\bar{a}}{\bar{a}_{max}}\right)^{-3(1+\omega)}\left(\frac{R}{R_{max}}\right)^3\right] \nonumber\\
\label{VT2}
\end{eqnarray}
where $\omega$ is the equation of state of dark energy and $q=\left(\frac{\rho_{DE}}{\rho_{DM}}\right)_{t=t_{max}}$, which is the ratio of energy densities of dark energy and dark matter inside the spherical over-dense regions at the turnaround time $t= t_{max}$. Here and throughout the paper, $a$ and $\bar{a}$ represent the scale factor of the spherical over-dense region and the background, respectively. Since the physical radius $R = r a(t)$, at the turnaround time, when the over-dense region reaches its maximum physical radius, the scale factor of the over-dense region also reaches its upper-limit $a_{max}$, and at that moment, the scale factor of background is $\bar{a}_{max}$. At the virialization time $t = t_{vir}$, the scale factors of the over-dense region and background become $a_{vir}$ and $\bar{a}_{vir}$, respectively. Using Eq.~(\ref{VT2}) and the virialization condition $\left(V_T + \frac12 R \frac{\partial V_T}{\partial R}\right)_{t=t_{vir}}=(V_T)_{t=t_{max}}$, we can get the following cubic equation of $\eta$:
\begin{eqnarray}
4Q\eta^3 \left(\frac{\bar{a}_{vir}}{\bar{a}_{max}}\right)^{-3(1+\omega)} -2 \eta (1+Q) +1 =0\,\, ,
\end{eqnarray}
where $Q=-(1+3\omega)\frac{q}{2}$.
It can be verified that for the vanishing value of $q$ (i.e., neglecting the dark energy effect), the solution of the above equation is $\eta=0.5$, which we obtained earlier for the top-hat collapse model. For a homogeneous cosmological constant model, where $\omega =-1$, the above cubic equation for $\eta$ becomes \cite{Lahav:1991wc, Shapiro}:
\begin{eqnarray}
4q\eta^3  -2 \eta (1+q) +1 =0\,\,.
\end{eqnarray}  
If we consider small value of $q$, the solution of the above equation for $\eta$ can be written as \cite{Shapiro}
\begin{equation}
\eta= 0.5 - 0.25 q -0.125 q^2+ \mathcal{O}(q^3)\,\, ,
\label{etaDE}
\end{equation}
which implies the value of $\eta$ is always less than $0.5$ for models involving the cosmological constant. The presence of $\Lambda$  makes the over-dense regions collapse more to attain the virialization state. 

In the homogeneous dark energy model, since the background universe continues expanding after the virialization of the over-dense regions, the density of the dark energy (with $\omega\neq -1$) in the virialized over-dense region also changes with time, and this is a big problem with the homogeneous dark energy model. This problem is discussed elaborately in \cite{Maor:2005hq}. This problem does not appear for models involving the cosmological constant since the density of dark energy always remains constant in such cases. 
The aforementioned problem is resolved in clustered dark energy models, where at the galaxy cluster scale, dark energy can cluster and virialize inside the over-dense regions. In this scenario, the total gravitational potential energy of the spherical over-dense regions can be written as \cite{Maor:2005hq}:
\begin{widetext}
\begin{eqnarray}
V_T = -\frac{3M^2}{5R}-(2+3\omega)\frac{3M^2}{5R}q\left(\frac{R}{R_{max}}\right)^{-3\omega}-(1+3\omega)\frac{3M^2}{5R}q^2\left(\frac{R}{R_{max}}\right)^{-6\omega}\,\,  ,
\end{eqnarray}
where each of the integrations in Eq.~(\ref{VT1}) has non-zero value. Using the virialization condition and the above expression of the total gravitational potential energy one can get the following equation for $\eta$ \cite{Maor:2005hq}
\begin{eqnarray}
\left[1+(2+3\omega)q+(1+3\omega)q^2\right]\eta-\frac12(2+3\omega)(1-3\omega)q\eta^{-3\omega}-\frac12(1-6\omega)(1+3\omega)q^2\eta^{-6\omega}=\frac12\,\, .
\end{eqnarray} 
\end{widetext} 
There exists another scenario where the dark energy only can cluster inside the spherical over-densities; however, it cannot virialize at that scale. For this scenario, the total potential energy of the spherical regions can be written as
\begin{eqnarray}
V_T=-\frac{3M^2}{5R}\left[1+q\left(\frac{R}{R_{max}}\right)^{-3\omega}\right]\,\, ,
\end{eqnarray}
from which we get the following equation for $\eta$ \cite{Maor:2005hq}
\begin{eqnarray}
\eta(1+q)-\frac{q}{2}(1-3\omega)\eta^{-3\omega}=\frac12\,\, .
\end{eqnarray} 
Fig.~(\ref{etavsq1}) depicts  how much the value of $\eta$ deviates from $0.5$ for different values of $q$ if we do not neglect the dark energy effect in the evolution of the spherical over-dense regions. In that figure, the brown line shows how $\eta$ changes with $q$ in those scenarios where dark energy can cluster and virialize inside the over-dense regions of dark matter. For this scenario, one can verify that $\eta$ is always greater than $0.5$. However, in the case where clustered dark energy cannot virialize, the virialized radius of the spherical over-dense region becomes smaller than half of the turnaround radius (i.e., $\eta < 0.5$) which is shown by the green line in Fig.~(\ref{etavsq1}). For both these cases, the equation of the state of dark energy is $\omega =-0.75$. On the other hand, the blue curve in Fig.~(\ref{etavsq1}) shows $\eta <0.5$ for the case involving the cosmological constant, however, the value of $\eta$ in this scenario is greater than that in the scenario where dark energy can cluster but cannot virialize. Therefore, we can see that the presence of negative pressure in the dark energy fluid can create distinguishable large-scale structures of dark matter.

In the next section, we use a two-fluid model to describe one of the cosmological scenarios discussed above, where the dark energy is homogeneous and it influences the collapsing dynamics of the over-dense dark matter region.
\begin{figure*}
\centering
\begin{minipage}[b]{.4\textwidth}
\centering
\includegraphics[width=2.8in,height=2.8in]{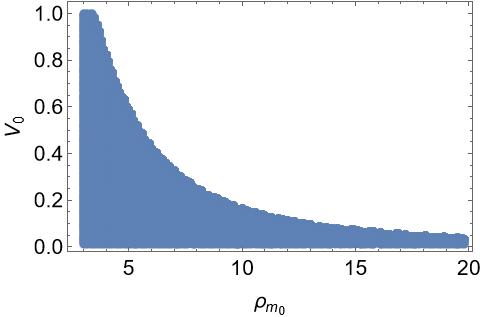}
\caption{Figure depicts the allowed parameters' space (i.e., shown by blue shaded region) of $V_0$ and $\rho_{m_0}$ for which the over-dense region collapses in the presence of quintessence-like scalar field after reaching its maximum physical radius.}
\label{regionplot1}
\end{minipage}\qquad
\begin{minipage}[b]{.4\textwidth}
\centering
\includegraphics[width=2.8in,height=2.8in]{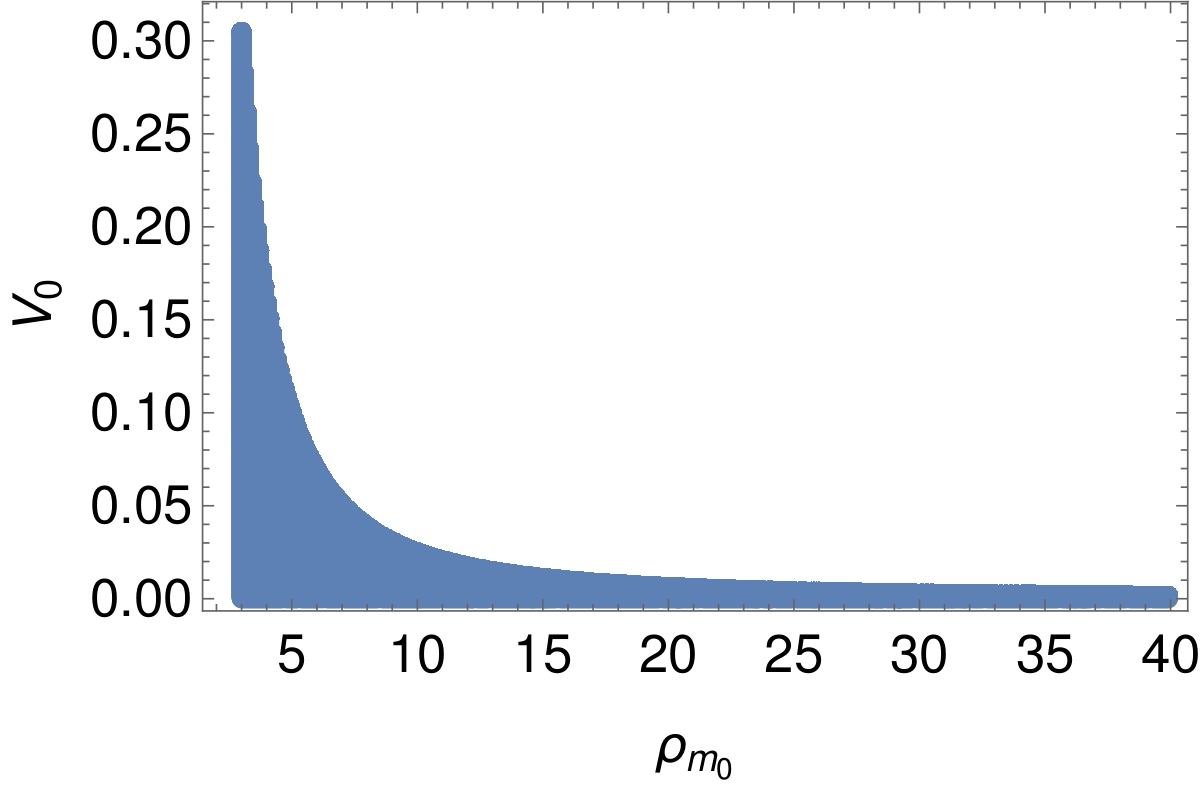}
\caption{Figure depicts the allowed parameters' space (i.e., shown by blue shaded region) of $V_0$ and $\rho_{m_0}$ for which the over-dense region collapses in the presence of phantom-like scalar field after reaching its maximum physical radius.}
\label{regionplot2}
\end{minipage}
\end{figure*}

\section{Gravitational collapse in the presence of dust-like matter and a scalar field}
\label{sec3}
As we discussed in the previous section, in this paper, we study the dynamics of a perfect fluid made of dust-like matter and a scalar field ($\phi(t)$) in order to understand the structure formation of dark matter in the presence of dark energy. Since we consider a minimally coupled scalar field with the dust-like matter, the energy-momentum tensor of the resultant fluid ($T^{\mu\nu}$) can be written as the sum of the energy-momentum tensors of the scalar field and the matter: 
\begin{eqnarray}
T^{\mu\nu}=(T^{\mu\nu})_{m}+(T^{\mu\nu})_{\phi}\,\, ,
\end{eqnarray}
where $(T^{\mu\nu})_{m}$ and $(T^{\mu\nu})_{\phi}$ correspond to the energy-momentum tensor of dust-like matter and the scalar field respectively. Therefore, $(T_{\nu}^{\mu})_{\phi}=\rm{diag}(-\rho_{\phi},p_{\phi},p_{\phi},p_{\phi})$ and $(T_{\nu}^{\mu})_m=\rm{diag}(-\rho_{m},0,0,0).$
Here, we consider the collapsing fluid is homogeneous and spherically symmetric. In order to model the dynamics of the over-dense region of dark matter in the presence of dark energy, we use closed FLRW spacetime:
\begin{eqnarray} \label{3}
ds^{2}=-dt^{2}+\frac{a^{2}(t)}{1-kr^{2}}dr^{2}+r^{2}a^{2}(t)(d\theta^{2}+\sin^{2}\theta d\Phi^{2})\,\,,\nonumber\\ 
\end{eqnarray}
where $a(t)$ is the scale factor of the over-dense region and the constant $k$ can be $0,\pm1$. If $k=0$ we have a flat spatial part whereas negative and positive $k$ imply an open or closed spatial section. We could have taken the metric of the over-dense region as a spatially flat FLRW metric, in that case, there will be no turnaround radius. We want to generalize the top-hat collapse in the presence of dark energy and for this, we require a turnaround. For a continual gravitational collapse, singularity forms when the scale factor $a(t)$ becomes zero at a comoving time $t_s$. At the initial stage of the gravitational collapse ($t=0$), $a(t)$ can attain any positive definite value that can always be rescaled to one. Therefore, we consider $a(t=0) = 1$. Since dark matter and dark energy should also be present in the background of the over-dense regions, we model the background by the above-mentioned two-fluid model, and we describe the dynamics of the background using flat FLRW spacetime:
\begin{eqnarray} \label{flatFLRW}
ds^{2}=-dt^{2}+\bar{a}^{2}(t)dr^{2}+r^{2}\bar{a}^{2}(t)(d\theta^{2}+\sin^{2}\theta d\Phi^{2})\,\,,\nonumber\\ 
\end{eqnarray}
where as mentioned before, the scale factor of the background is denoted by $\bar{a}(t)$.

In the present paper, as stated above, we describe the dynamics of the over-dense region by closed FLRW space-time, and the background is modeled by flat FLRW space-time. However, in order to describe a matter flux through the boundary of the over-dense region, in the immediate neighborhood of the over-dense region, we consider an external generalized Vaidya space-time. It should be noted that we do not consider Vaidya space-time as a background space-time. The background at the Hubble scale is modeled by flat FLRW space-time. Vaidya space-time is used only to describe the local dynamics of matter around the boundary of the over-dense patches.

The boundary of the over-dense region is a timelike hyper-surface $\Sigma = r-r_b = 0$, and 
the dynamical spacetime structure that we consider here is internally ($\mathcal{V}^{-}$) closed FLRW metric and externally ($\mathcal{V}^{+}$) exploding generalized Vaidya spacetime:
\begin{eqnarray}
dS^2_{-}&=&-dt^2+a^2(t)\left(\frac{dr^2}{1-r^2}+r^2d\Omega^2\right)\,\,\nonumber\\
&=& -dt^2 + a^2(t) d\Psi^2 +a^2(t)\sin^2\Psi d\Omega^2\,\, ,\\
\nonumber\\
dS^2_{+}&=& -\left(1-\frac{2M(r_v , v)}{r_v}\right)dv^2 - 2dv dr_v + r_v^2 d\Omega^2\,\, , \nonumber\\
\end{eqnarray}
where we consider co-moving radius $r=\sin\Psi$ and $r_v$ and $v$ are the coordinates corresponding to the generalised Vaidya spacetime. At the timelike hyper-surface ($\Sigma$) where the internal and external spacetimes match with each other, $\Psi$ becomes $\Psi_b$ and the $v$ and $r_v$ become the function of co-moving time $t$. Therefore, at $\Sigma$, we can write down the induced metric from both the sides as,
\begin{eqnarray}
dS^2_{-}|_{\Sigma}&=& -dt^2 + a^2(t)\sin^2\Psi_b d\Omega^2\,\, ,\\
\nonumber\\
dS^2_{+}|_{\Sigma}&=& -\left(\dot{v}^2-\frac{2M(r_v , v)}{r_v}\dot{v}^2+2\dot{v}\dot{r}_v\right)dt^2+ r_v^2 d\Omega^2\,\, , \nonumber\\
\end{eqnarray}
where $\dot{v}$ and $\dot{r}_v$ are the partial derivatives of $v$ and $r_v$ with respect to co-moving time $t$. As we know, for the smooth matching of two spacetimes at a hyper-surface, the necessary and sufficient condition is that the induced metric ($h_{ab}$) and the extrinsic curvature ($K_{ab}$) from both the sides should match at the junction. From the induced metric matching of the above spacetime structures on $\Sigma$ yields:
\begin{eqnarray}
\left(\dot{v}^2-\frac{2M(r_v , v)}{r_v}\dot{v}^2+2\dot{v}\dot{r}_v\right) &=& 1\,\, ,\\
r_v &=& a(t)\sin\Psi_b\,\, .
\end{eqnarray}
\begin{figure*}
\subfigure[Variation of $a$ with time]
{\includegraphics[width=81mm,height=52mm]{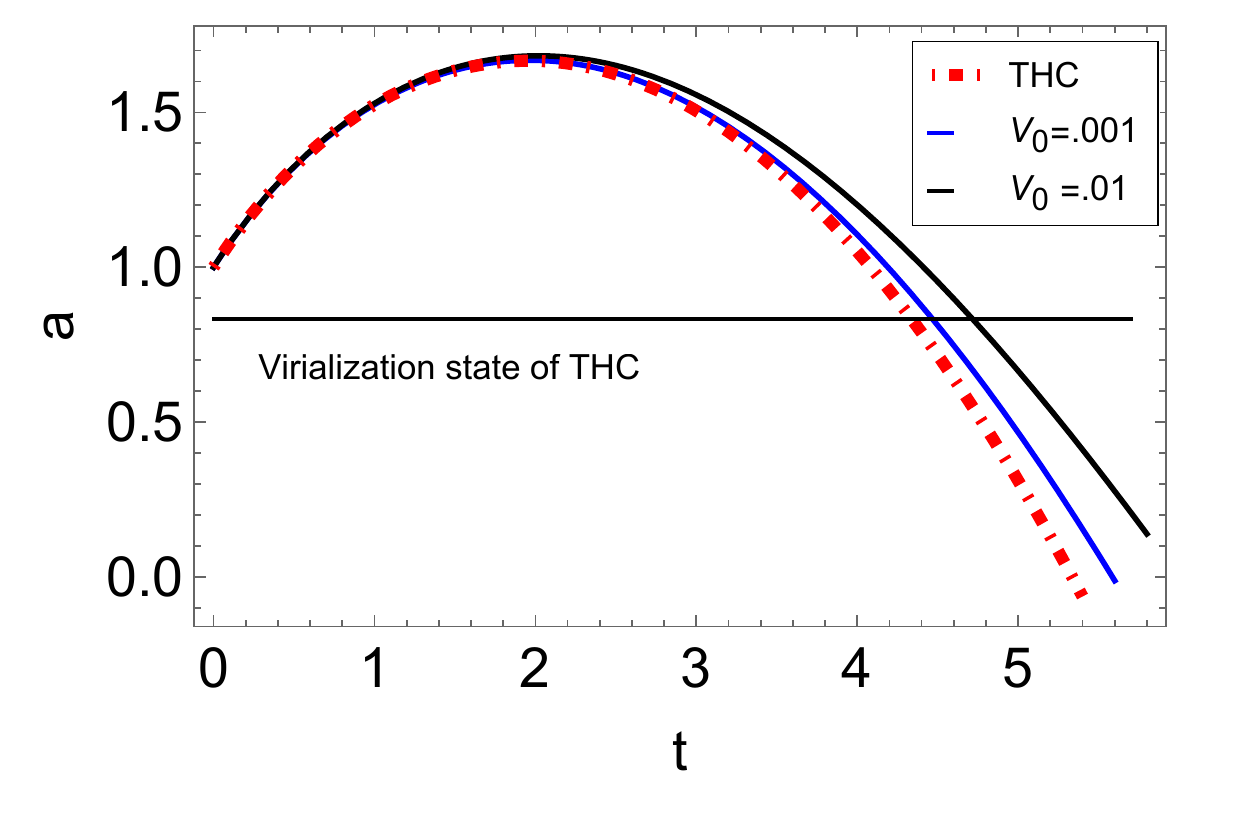}\label{Firsta}}
\hspace{0.2cm}
\subfigure[Variation of $\omega_{\phi}$ with time]
{\includegraphics[width=82mm,height=50mm]{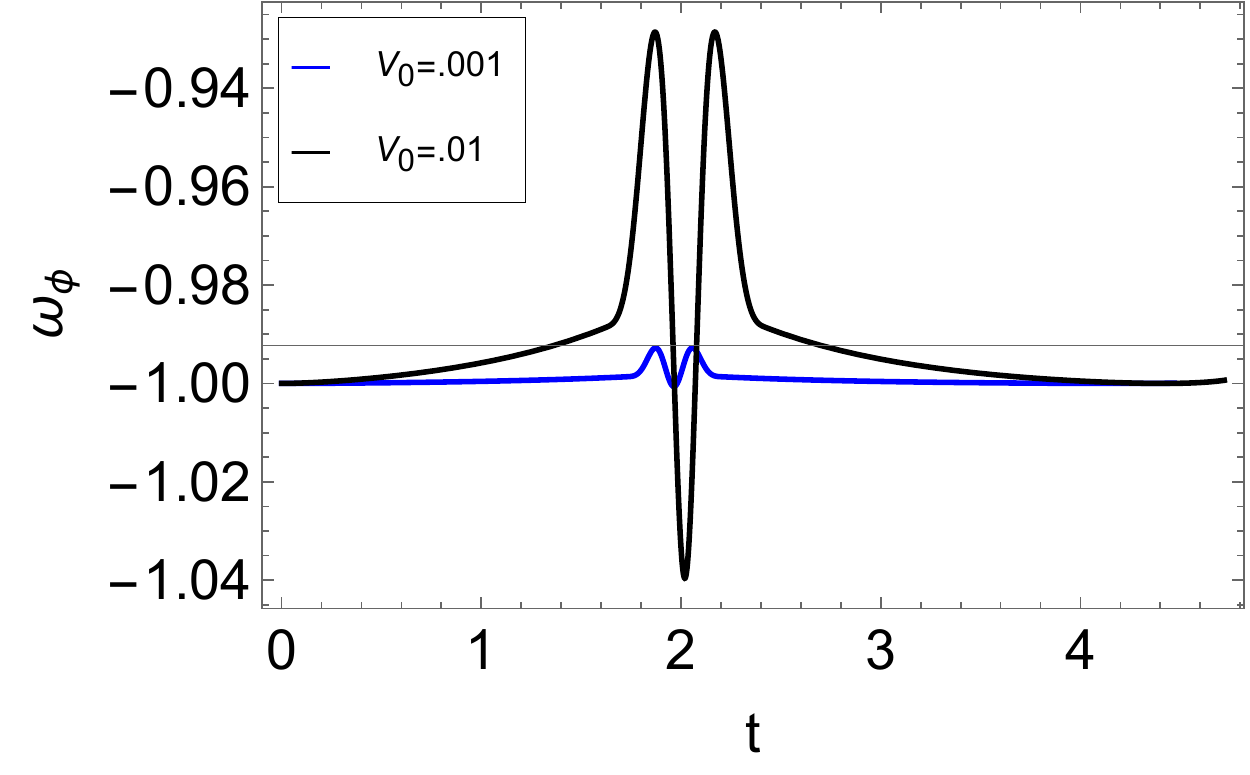}\label{Firstb}}
\hspace{0.2cm}
\subfigure[Variation of $\frac{\rho_{\phi}}{\Bar{\rho}_{\phi}}$ with time]
{\includegraphics[width=82mm,height=50mm]{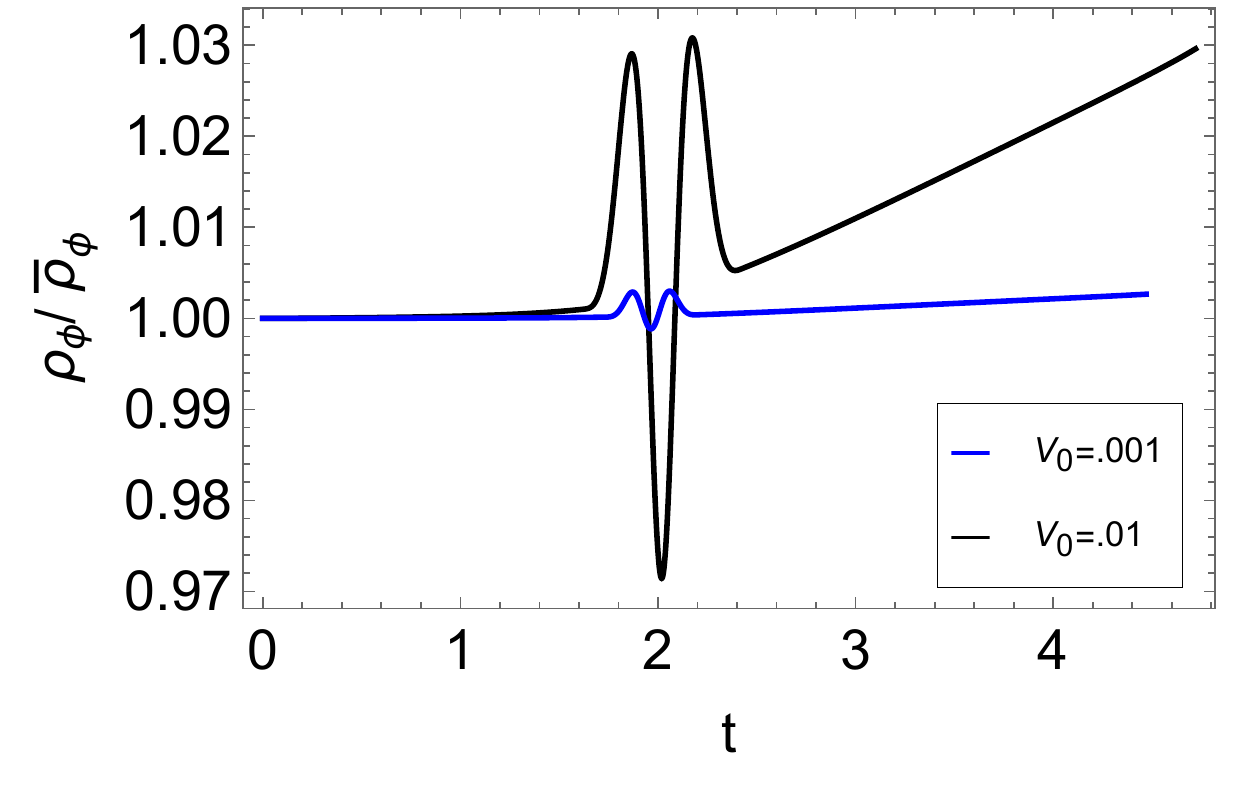}\label{Firstc}}
\hspace{0.2cm}
\subfigure[Variation of $\omega_t$ with time]
{\includegraphics[width=85mm,height=54mm]{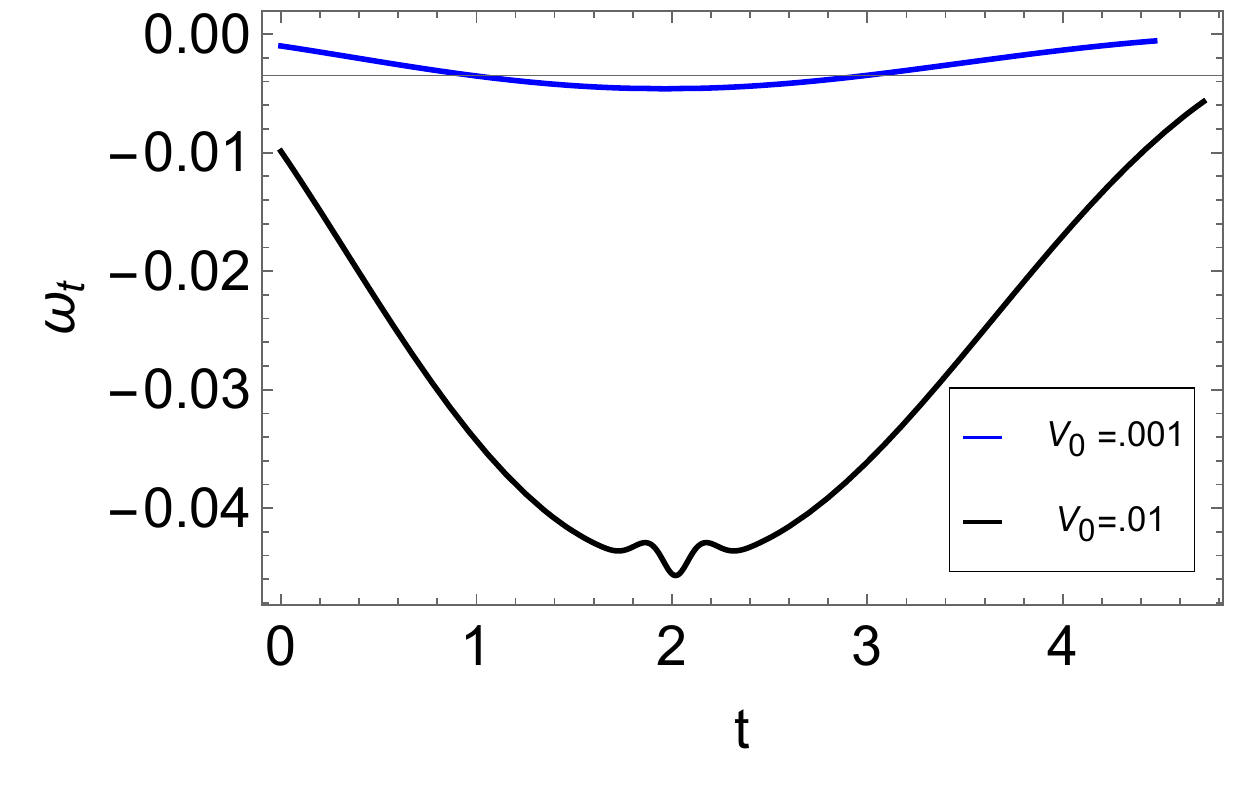}\label{Firstd}}
\hspace{0.2cm}
\caption{Figure shows variation of different variables with variation of $V_{0}$ for scalar potential $V(\phi)=V_{0}e^{-\lambda\phi}$ for Quintessence field.}
\label{First}
\end{figure*}
\begin{figure*}
\subfigure[Variation of $a$ with time]
{\includegraphics[width=81mm,height=52mm]{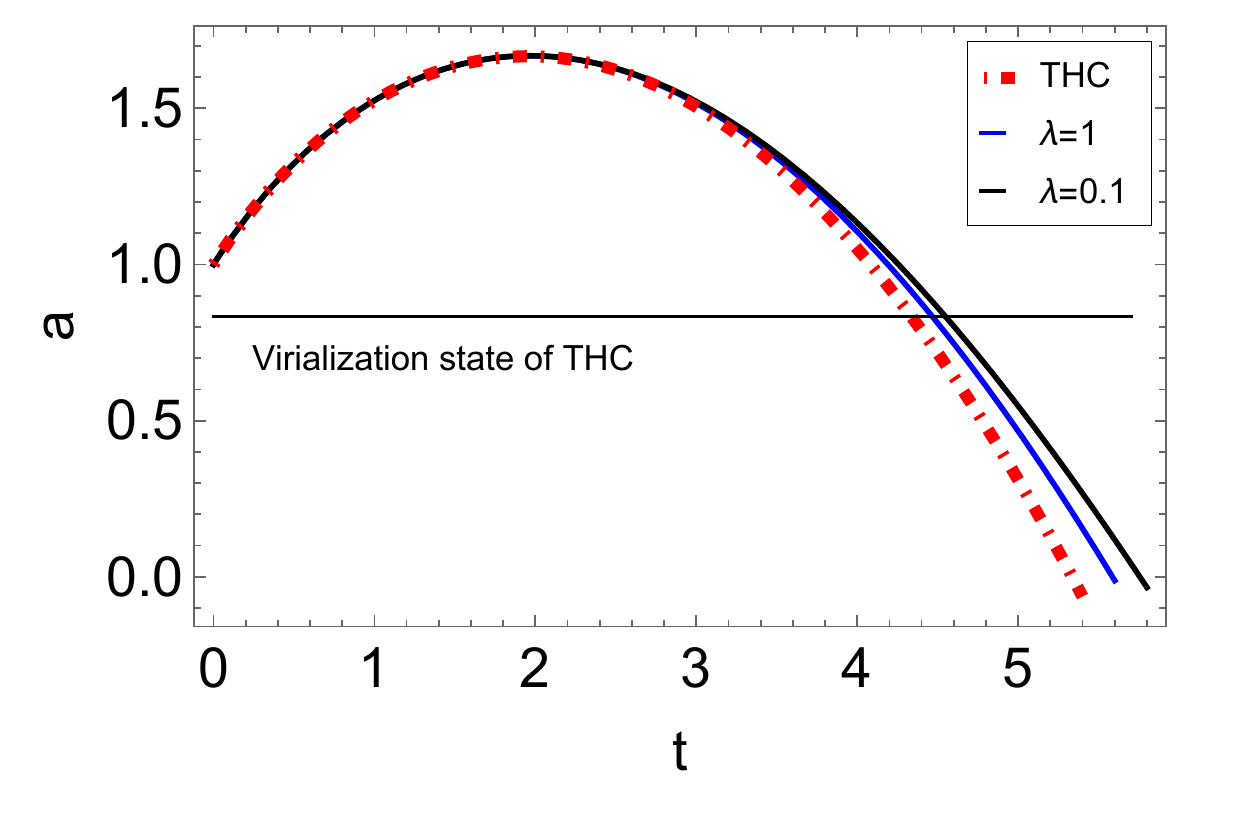}\label{Thirda}}
\hspace{0.2cm}
\subfigure[Variation of $\omega_{\phi}$ with time]
{\includegraphics[width=82mm,height=50mm]{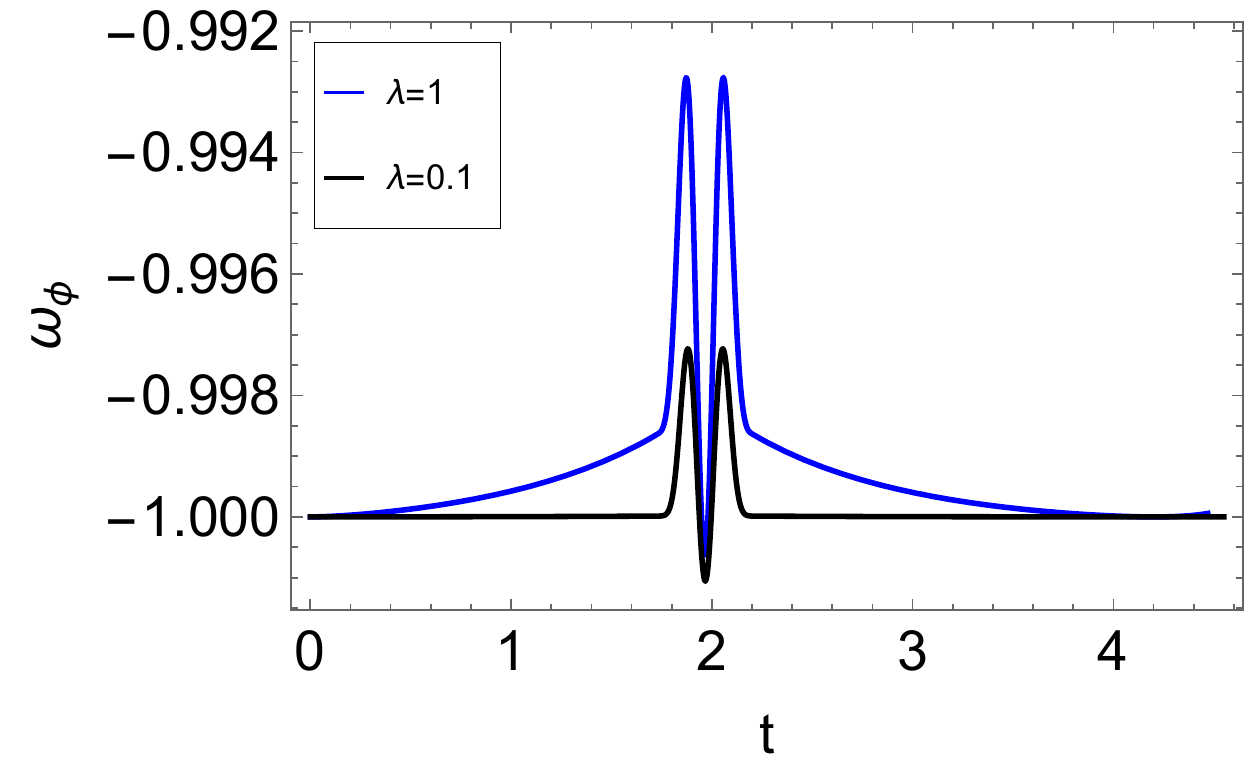}\label{Thirdb}}
\hspace{0.2cm}
\subfigure[Variation of $\frac{\rho_{\phi}}{\Bar{\rho}_{\phi}}$ with time]
{\includegraphics[width=82mm,height=50mm]{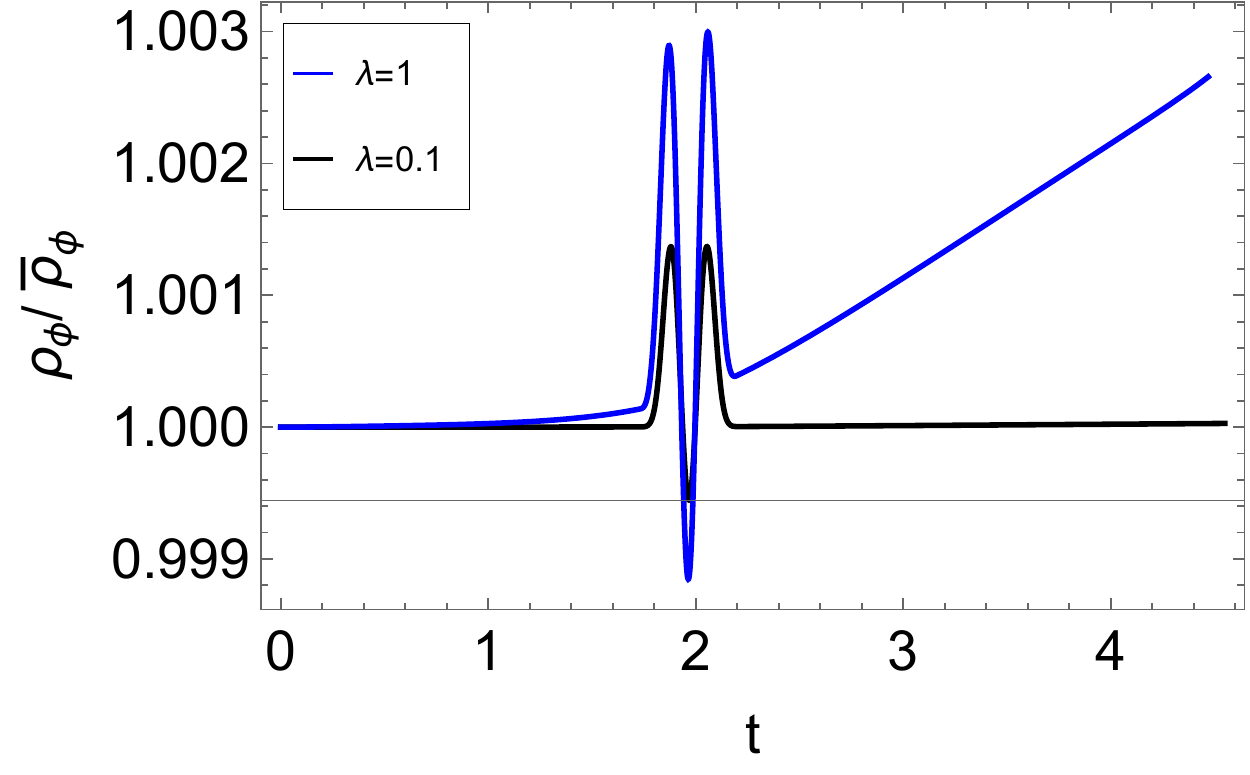}\label{Thirdc}}
\hspace{0.2cm}
\subfigure[Variation of $\omega_t$ with time]
{\includegraphics[width=85mm,height=54mm]{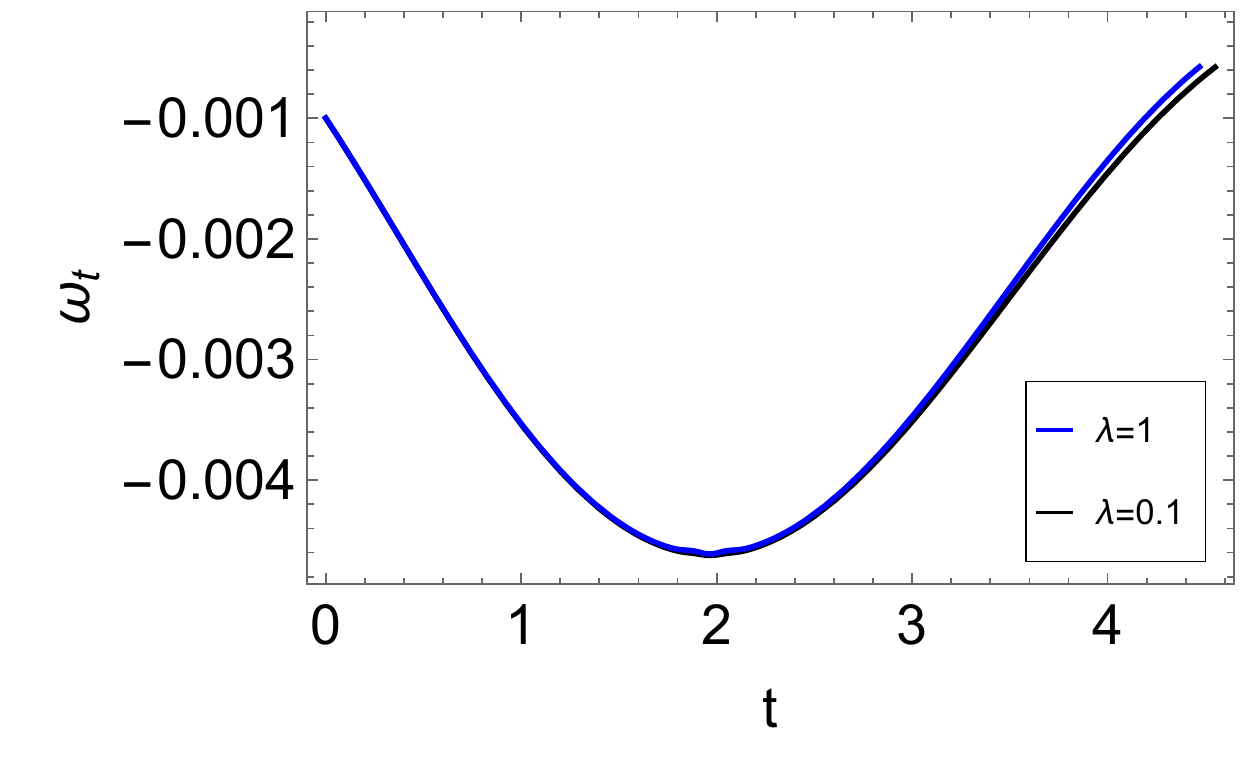}\label{Thirdd}}
\hspace{0.2cm}
\caption{Figure shows variation of different variables with variation of $\lambda$ for scalar potential $V(\phi)=V_{0}e^{-\lambda\phi}$ for Quintessence field.}
\label{Third}
\end{figure*}

In order to calculate the extrinsic curvature ($K_{ab}$), one needs the information of the spacelike normal ($n^{\alpha}$) to  $\Sigma$ from both the sides. From the side $\mathcal{V}^{-}$, the four velocity ($u^{\alpha}$) of the comoving shell $\Sigma$ can be written as: $u^{\alpha}_{-}\equiv\lbrace 1,0,0,0\rbrace$. Using $(n_{\alpha})_{-}n^{\alpha}_{-}=1$ and $(n_{\alpha})_{-}u^{\alpha}_{-}=0$, we get $$(n_{\alpha})_{-}\equiv \lbrace 0,a(t),0,0\rbrace .$$
For $\mathcal{V}^{+}$, we can write down the following expression of $u^{\alpha}_{+}, n^{\alpha}_{+}$ as,
\begin{eqnarray}
u^{\alpha}_{+}&\equiv& \lbrace\dot{v},\dot{r}_v , 0, 0\rbrace\,\, ,\\
n^{\alpha}_{+}&\equiv& \lbrace-\frac{1}{\sqrt{1-\frac{2M}{r_v}+2\frac{dr_v}{dv}}}, \frac{1-\frac{2M}{r_v}+\frac{dr_v}{dv}}{\sqrt{1-\frac{2M}{r_v}+2\frac{dr_v}{dv}}},0,0\rbrace\,\, .\nonumber\\
\end{eqnarray}
Using the expressions of $u^{\alpha}$ and $n^{\alpha}$ from both the sides, we get the following expressions of azimuthal components of extrinsic curvature tensors: 
\begin{eqnarray}
K_{\theta\theta}^{-}&=& a(t)\sin\Psi_b \cos\Psi_b\,\, ,\\
K_{\theta\theta}^{+}&=& r_v \frac{1-\frac{2M}{r_v}+\frac{dr_v}{dv}}{\sqrt{1-\frac{2M}{r_v}+2\frac{dr_v}{dv}}}\,\, .
\end{eqnarray}
Equating $K_{\theta\theta}^+$ and $K_{\theta\theta}^-$, we get,
\begin{equation}
\cos\Psi_b = \frac{1-\frac{2M}{r_v}+\frac{dr_v}{dv}}{\sqrt{1-\frac{2M}{r_v}+2\frac{dr_v}{dv}}}\,.
\end{equation}
Equating the temporal components of $K_{tt}$ from both sides we get,
\begin{equation}
\label{Massderivative}
M(r_v,v)_{,r_v}=\frac{F}{2\sin\psi_b a(t)}+\sin^2\psi_b a\ddot{a}\,\, ,
\end{equation} 
where $F$ is the Misner-Sharp mass of the internal collapsing spacetime which should follow the following condition at the boundary,
\begin{equation}
F(t,\sin\psi_b)=2M(r_v,v)\,\,.
\end{equation}
From Eq.~(\ref{Massderivative}), it can be seen how the flux of the matter at the boundary depends upon the scale factor and the Misner-Sharp mass ($F$) of the collapsing spacetime. In the present case, $F$ is a function of time only, since it represents the internal homogeneous two-fluid system. Due to the time dependence of $F$, pressure is non-zero internally and it can be written as:
\begin{eqnarray}
    p = -\frac{\dot{F}}{\dot{R}R^2}\,\, .
\end{eqnarray}
A non-zero pressure at the boundary of a system implies the existence of non-zero matter flux through the boundary and that is the very reason why we consider generalized Vaidya spacetime in the immediate neighborhood of the internal two-fluid system. From the above expression of pressure, it can be understood that the presence of negative pressure at the boundary of an internal spacetime implies an inward matter flux through the boundary for an expanding scenario and an outward matter flux for a collapsing scenario. In our model, the non-zero internal pressure is generated due to the presence of a scalar field, and therefore, the scalar field is responsible for the non-zero flux through the boundary. On the other hand, the matter-field part of the two-fluid system does not leak out of the boundary, since it has zero pressure. Only the scalar field continuously is leaking out/in throughout the whole dynamics of the over-dense region. If there exists a non-minimal coupling between the matter and scalar field then a non-zero pressure at the boundary can make the matter-field flux out/in along with the scalar field. In this paper, we consider only the minimal coupling between the matter and scalar field, and therefore, the above-mentioned scenario where the matter has non-zero flux at the boundary is not possible. The flux of the scalar field from inside gives rise to non-zero components of the energy-momentum tensor of the external generalized Vaidya spacetime which is seeded by a fluid composed of null dust and perfect fluid. Therefore, the energy-momentum tensor of the internal spacetime and the external spacetime can be respectively written as:
\begin{eqnarray}
    T_{\mu\nu}^- &=& (\rho_m+\rho_\phi+p_\phi)u_\mu^-u_\nu^- + p_\phi g_{\mu\nu}^-\,\, ,\nonumber\\
    T_{\mu\nu}^+ &=& \bar{\epsilon} l_\mu k_\nu + (\epsilon+\mathcal{P})\left(l_{\mu}k_{\nu}+l_{\nu}k_{\mu}\right)+\mathcal{P}g_{\mu\nu}^+\,\, ,
\end{eqnarray}
where $\bar{\epsilon}$, $\epsilon$, and $\mathcal{P}$ can be written as:
\begin{eqnarray}
    \bar{\epsilon}=-\frac{2 M,v}{r_v^2}, \hspace{0.2cm} \epsilon=\frac{2M,_{r_v}}{r_v^2}, \hspace{0.2cm} \textrm{and} \hspace{0.2cm} \mathcal{P}=-\frac{M,_{r_v r_v}}{r_v},
\end{eqnarray}
and $l^\mu, k^\mu$ are two null vectors which follow the condition: $l^\mu k_\mu = -1$. Due to the existence of non-zero pressure at the boundary, the flux from the internal spacetime at the boundary seeds the components of the energy-momentum tensor of the external generalized Vaidya spacetime. In the next section, we show the dynamics of the two-fluid system by solving Einstein's equations for the internal spacetime. Using the freedom to choose one free function, we consider the scalar field is either a quintessence field or a phantom field, and since the matter is minimally coupled with the scalar field, $\rho_m$ varies as $\frac{1}{a^3}$. This prior consideration makes the matter part evolve like a closed dust ball, while the internal density of the scalar field stays almost constant throughout the evolution which implies a non-zero flux of the scalar field through the boundary. We consider the initial matter density $\rho_{m_0}$ to be $10^3 - 10^4$ times greater than the initial density of the scalar field which allows us to use the virialization technique discussed in Sec.~(\ref{sec2}) in order to understand the virialization process of the two-fluid system, though the two-fluid system in our model is not a closed system.     
\section{Gravitational collapse solutions of matter in the presence of Quintessence and Phantom-like scalar fields}
\label{sec4}

Using Einstein's equation for the FLRW space-time (Eq.~(\ref{3},\ref{flatFLRW})), one can write down the effective density and pressure of the resultant fluid as,
\begin{eqnarray} \label{8}
\rho&=&\rho_{\phi}+\rho_{m}=\frac{1}{2}\epsilon\dot{\phi}^{2}+V(\phi)+\rho_{m}=\frac{3\dot{a}^{2}}{a^{2}}+\frac{3k}{a^{2}}\,\, ,\\
p&=&p_{\phi}=\frac{1}{2}\epsilon\dot{\phi}^{2}-V(\phi)=-\frac{2\ddot{a}}{a}-\frac{\dot{a^{2}}}{a^{2}}-\frac{k}{a^{2}}\,\, ,\label{9}
\end{eqnarray}
where the $V(\phi)$ is the potential of the scalar field, $\epsilon$ is a real-valued constant, $k$ represents the curvature of 3-space, and over-dot denotes the time derivatives of the function. In this section, the above expressions of $\rho$ and $p$ and all other differential equations are written in a general way, where $k=0$ implies the corresponding equations are related to the background,  on the other hand, $k=1$ implies they are related to the over-dense region. 
From Eq.~(\ref{8}) and Eq.~(\ref{9}), it can be seen that there are four unknown functions: $V(\phi), \phi(a), \dot{a}(a)$ and $\rho_m(a)$ and two differential equations, and therefore, we have the freedom to choose two free functions along with the initial conditions to solve the differential equations. As stated before, here we consider the scenario where the scalar field is minimally coupled with dust-like matter. Therefore, the energy-momentum tensors of matter and scalar field follow the conservation equation separately: 
\begin{eqnarray}
    \nabla_a T^{ab}_{\phi} &=& 0 \implies \ddot{\phi}+ 3\frac{\dot{a}}{a} \dot{\phi} + V_{,\phi} = 0\,\, ,\\
    \nabla_a T^{ab}_{m} &=& 0 \implies \dot{\rho}_m + 3\frac{\dot{a}}{a}\rho_m = 0\,\, .
\end{eqnarray}
Consequently we have $\rho_m\propto \frac{1}{a^3}$. This shows that ultimately we have to choose only one function out of  $V(\phi), \phi(a), \dot{a}(a)$ to solve the differential Eqs.~(\ref{8}),(\ref{9}). 

Using the expression of energy density and pressure of the scalar field, we can write
\begin{equation} \label{10}
\rho_{\phi}+p_{\phi}=\epsilon\dot{\phi^{2}}=\epsilon\phi_{,a}^{2}\dot{a}^{2}\,\, ,
\end{equation}
where we use the chain rule $\dot{\phi^{2}}=\phi_{,a}^{2}\dot{a}^{2}$
where $\phi_{,a}$ imply a derivative with respect to $a$. From Eq.~(\ref{8}) we get
\begin{equation} \label{11}
\dot{a}=\pm\sqrt{\frac{\rho_{\phi}+\rho_{m}}{3}a^{2}-k}\,\, ,
\end{equation}
where the $+$ and $-$ signs are for expanding and collapsing scenarios, respectively.
Now, differentiating (\ref{11}) with respect to the comoving time ($t$) we get,
\begin{equation} \label{12}
\ddot{a}=\frac{a}{3}\left[\rho_{\phi}+\rho_{m}+\frac{a}{2}(\rho_{\phi,a}+\rho_{m,a})\right]\,\, ,
\end{equation}
where $\rho_{\phi,a}$ and $\rho_{m,a}$ are derivatives of the scalar field energy density and the fluid energy density respectively, with respect to the scale factor $a$.

Using Eqs.~(\ref{10}) and (\ref{11}) we get
\begin{equation} \label{13}
\rho_{\phi}(1-\frac{\epsilon\phi_{,a}^{2}a^{2}}{3})-\rho_{m}\frac{\epsilon\phi_{,a}^{2}a^{2}}{3}+p_{\phi}+k\epsilon\phi_{,a}^{2}=0\,\, .
\end{equation}
From Eq.~(\ref{8}) we get
\begin{equation} \label{14}
p_{\phi}=\rho_{\phi}-2V(\phi)\,\, .
\end{equation}
Since quintessence like scalar field has positive kinetic energy, $\epsilon =1$ and we can write down the following expression of $\rho_{\phi}$
using Eqs.~(\ref{13}), (\ref{14})
\begin{equation} \label{16}
\rho_{\phi}=\frac{\frac{\rho_{m}\phi_{,a}^{2}a^{2}}{6}+V(\phi)-\frac{k\phi_{,a}^{2}}{2}}{(1-\frac{\phi_{,a}^{2}a^{2}}{6})}\,\, .
\end{equation}
Now, using Eqs.~(\ref{10}) ,(\ref{11}), (\ref{12}) and  (\ref{13}), we get
\begin{equation} \label{17}
\rho_{\phi,a}=\frac{-\phi_{,a}^{2}\rho_{\phi}a^{2}-(3+a^{2}\phi_{,a}^{2})\rho_{m}+3k\phi_{,a}^{2}}{a}-\rho_{m,a}\,\, .
\end{equation}
\begin{figure*}
\subfigure[Variation of $a$ with time for $\rho_{m_{0}}=5$]
{\includegraphics[width=81mm,height=52mm]{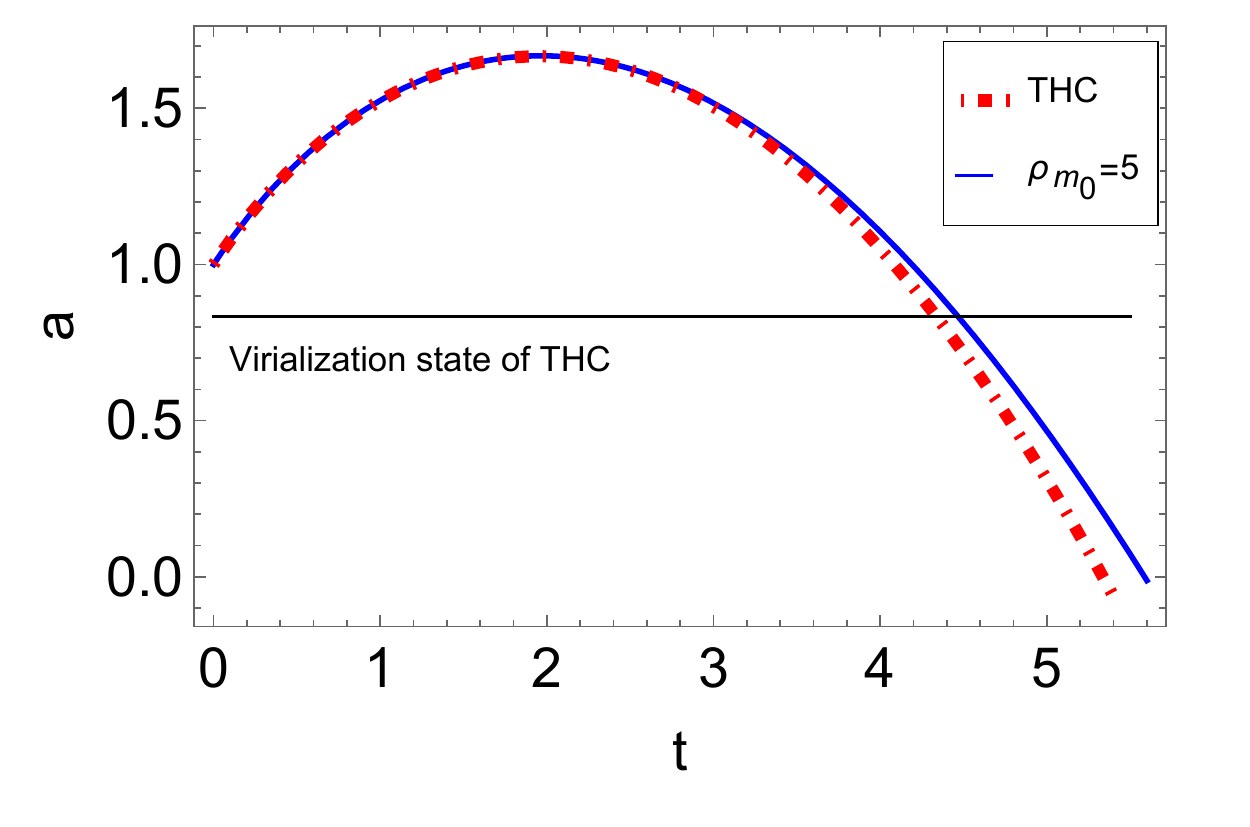}\label{Fourtha}}
\hspace{0.2cm}
\subfigure[Variation of $a$ with time for $\rho_{m_{0}}=6$]
{\includegraphics[width=81mm,height=52mm]{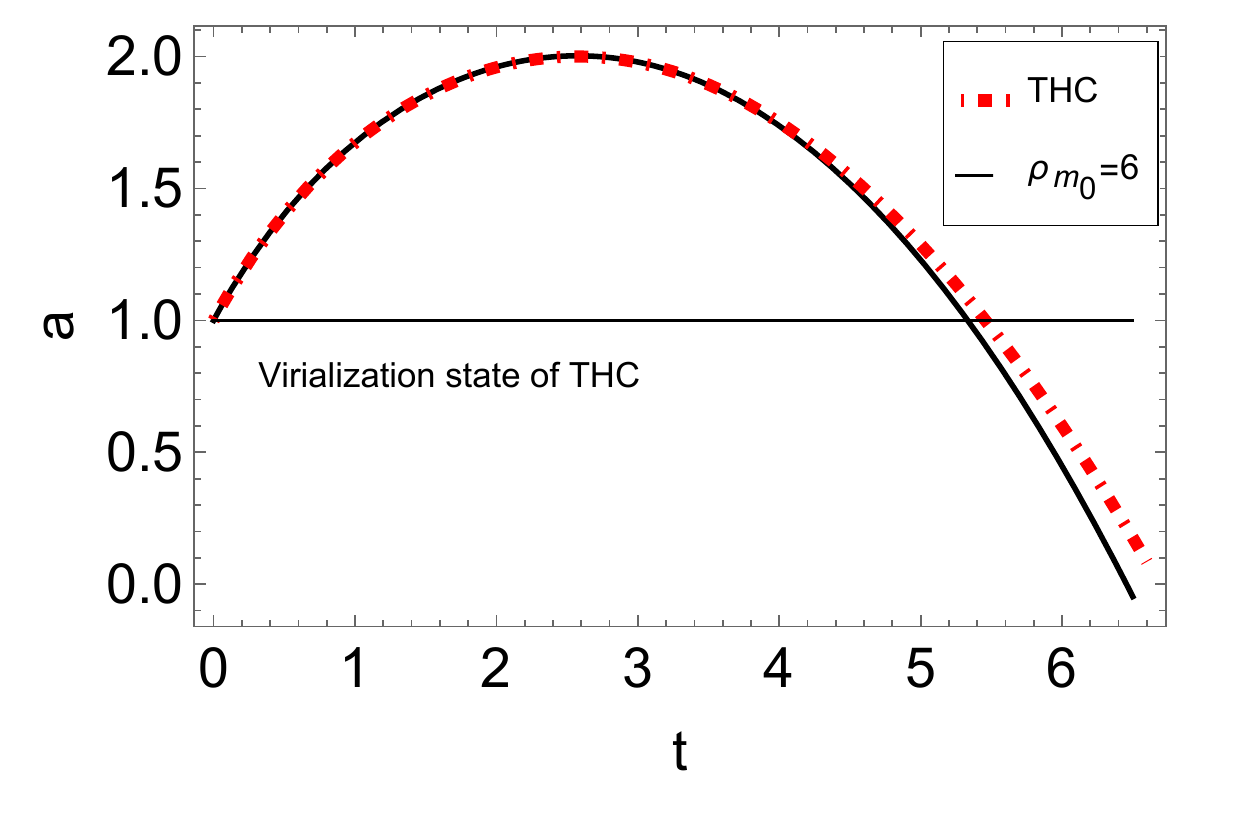}\label{Fourthb}}
\hspace{0.2cm}
\subfigure[Variation of $\omega_{\Phi}$ with time]
{\includegraphics[width=82mm,height=50mm]{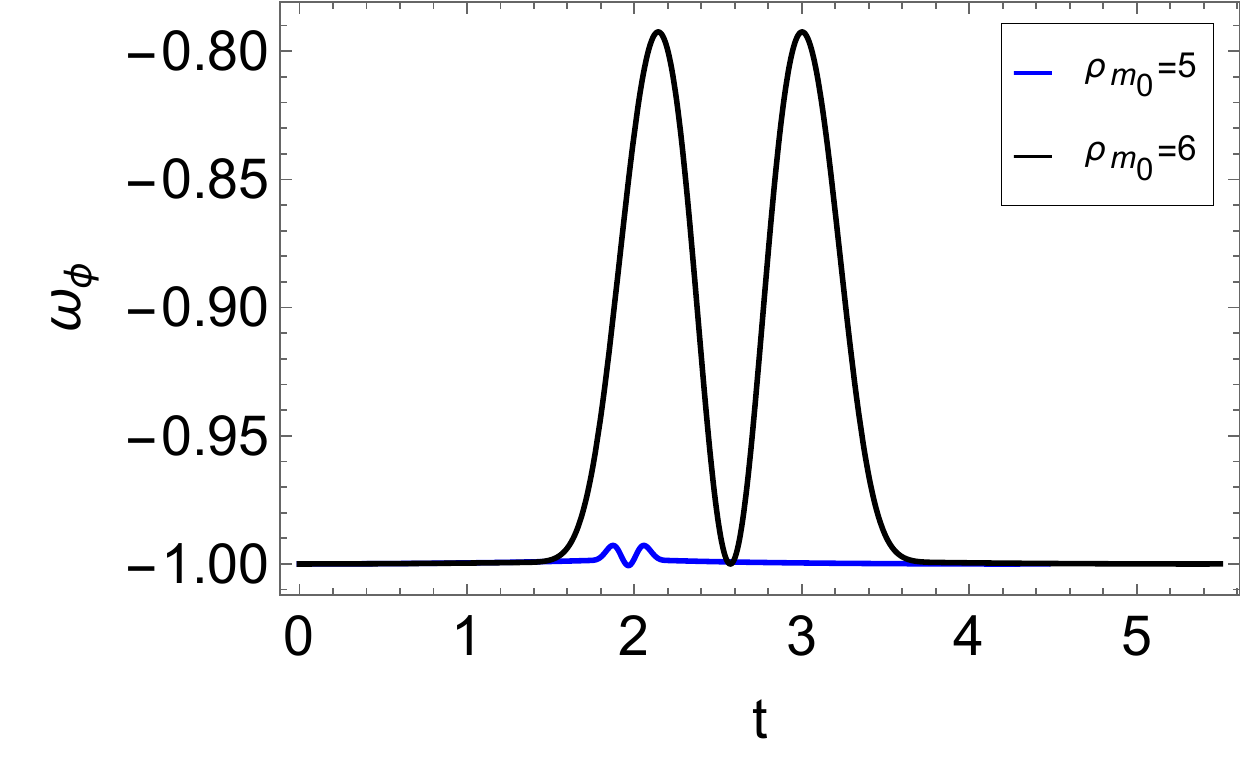}\label{Fourthc}}
\hspace{0.2cm}
\subfigure[Variation of $\frac{\rho_{\phi}}{\Bar{\rho}_{\phi}}$ with time]
{\includegraphics[width=82mm,height=50mm]{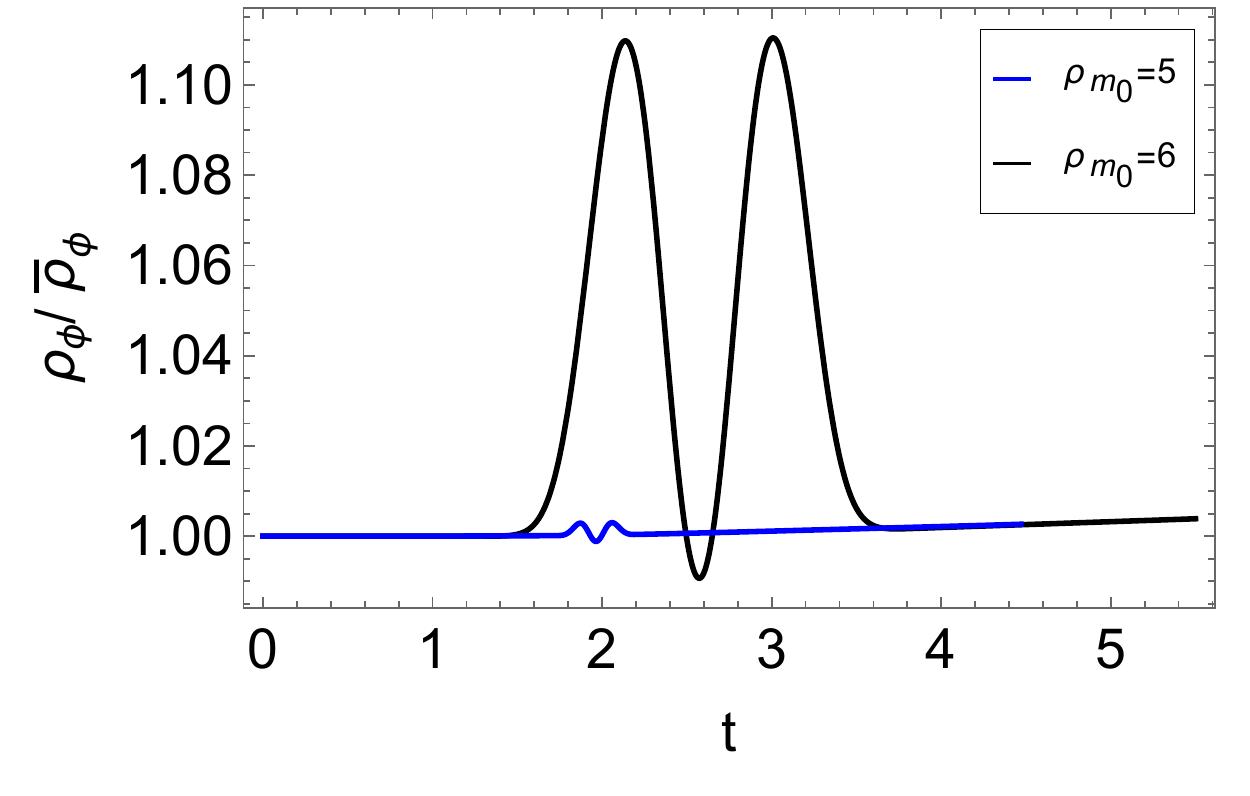}\label{Fourthd}}
\hspace{0.2cm}
\subfigure[Variation of $\omega_t$ with time]
{\includegraphics[width=85mm,height=54mm]{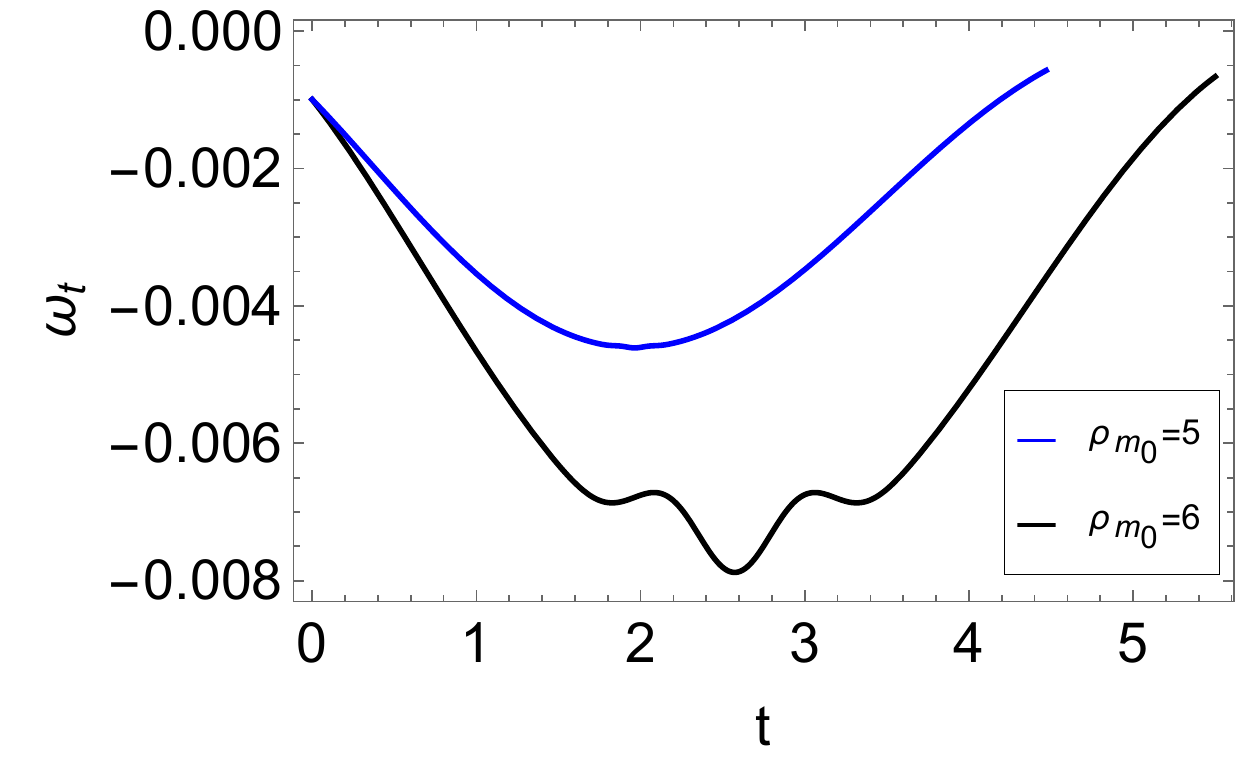}\label{Fourthe}}
\hspace{0.2cm}
\caption{Figure shows variation of different variables with variation of $\rho_{m_{_{0}}}$ for scalar potential $V(\phi)=V_{0}e^{-\lambda\phi}$ for Quintessence field.}
\label{Fourth}
\end{figure*}
Now, differentiating Eq.~(\ref{16}) with respect to $a$ and using equation Eq.~(\ref{17}) we get the following second order differential equation
\begin{widetext}
\begin{eqnarray}  \label{18}
\frac{-\phi_{,a}^{2}\rho_{\phi}a^{2}-(3+a^{2}\phi_{,a}^{2})\rho_{m}+3k\phi_{,a}^{2}}{a}-\rho_{m,a}&=&\frac{1}{3(1-\frac{\phi_{,a}^{2}a^{2}}{6})^{2}}\{3V_{,\phi}\phi_{,a}+\frac{\rho_{m,a}\phi_{,a}^{2}a^{2}}{2}+\rho_{m}a\phi_{,a}^{2}-\frac{\rho_{m,a}\phi_{,a}^{4}a^{4}}{12}\nonumber\\&+&\rho_{m}\phi_{,a}\phi_{,aa}a^{2}-\frac{V_{,\phi}\phi_{,a}^{3}a^{2}}{2}+V(\phi)a\phi_{,a}^{2}+V(\phi)a^{2}\phi_{,a}\phi_{,aa}-3k\phi_{,a}\phi_{,aa}-\frac{k\phi_{,a}^{4}a}{2}\}\,\, ,\nonumber\\
\end{eqnarray}
\end{widetext}
where $\phi_{,aa}$ is the second order derivative of scalar field with respect to $a$. As we have mentioned before, we have to choose only one function among $V(\phi), \phi(a), \dot{a}(a)$ to solve the dynamics of collapse. Therefore, here, we choose $V(\phi)=V_{0}e^{-\lambda\phi}$ which is generally considered as the potential of quintessence-like scalar fields \cite{Copeland:2006wr}.
Now, considering $V(\phi)=V_{0}e^{-\lambda\phi}$ , $\rho_{m}=\frac{\rho_{m_{0}}}{a^{3}}$ and $k=1$ we get
\begin{widetext}
\begin{eqnarray}  \label{19}
-4V_{0}e^{-\lambda\phi}\phi_{,a}a^{3}+9\phi_{,a}a &+&\frac{V_{0}e^{-\lambda\phi}\phi_{,a}^{3}a^{5}}{2}+\frac{\rho_{m_{0}}\phi_{,a}^{3}a^{2}}{4}-\phi_{,a}^{3}a^{3}+3\lambda a^{2}V_{0}e^{-\lambda\phi}-\frac{5\rho_{m_{0}}\phi_{,a}}{2}-\rho_{m_{0}}a\phi_{,aa}+3a^{2}\phi_{,aa}-V_{0}e^{-\lambda\phi}a^{4}\phi_{,aa}\nonumber\\&-&\frac{\lambda V_{0}e^{-\lambda\phi}\phi_{,a}^{2}a^{4}}{2}=0\,\, ,\nonumber\\
\end{eqnarray}
and for $k=0$,
\begin{eqnarray}  \label{20}
-4V_{0}e^{-\lambda\phi}\phi_{,a}a^{3}+\frac{V_{0}e^{-\lambda\phi}\phi_{,a}^{3}a^{5}}{2}+\frac{\rho_{m_{0}}\phi_{,a}^{3}a^{2}}{4}+3\lambda a^{2}V_{0}e^{-\lambda\phi}-\frac{5\rho_{m_{0}}\phi_{,a}}{2}-\rho_{m_{0}}a\phi_{,aa}-\frac{\lambda V_{0}e^{-\lambda\phi}\phi_{,a}^{2}a^{4}}{2}-V_{0}e^{-\lambda\phi}a^{4}\phi_{,aa}=0\,\, .\nonumber\\
\end{eqnarray}
\end{widetext}

We can now solve the above differential equations for $k=0, 1$ to get the functional form of $\phi(a)$ and using the solution of $\phi(a)$ and the differential Eqs.~(\ref{8}), (\ref{9}), we can get the expression of scale factor $a$ as a function of comoving time $t$. Since the differential Eq.~(\ref{19}) corresponds to $k=1$, solving that equation and using Eqs.~(\ref{8}), (\ref{9}), we can get the dynamics of the resultant fluid in the over-dense region. On the other hand, the solution of Eq.~(\ref{20}) shows the dynamics of the resultant fluid in the background, since that equation corresponds to $k=0$.

It is generally considered that the phantom-like scalar field has negative kinetic energy and therefore, for the phantom field $\epsilon = -1$. For the phantom field, the above two differential equations become,
\begin{widetext}
for $k=1$,
\begin{eqnarray}  \label{25}
4V_{0}e^{-\lambda\phi}\phi_{,a}a^{3}-9\phi_{,a}a&+&\frac{V_{0}e^{-\lambda\phi}\phi_{,a}^{3}a^{5}}{2}+\frac{\rho_{m_{0}}\phi_{,a}^{3}a^{2}}{4}-\phi_{,a}^{3}a^{3}+3\lambda a^{2}V_{0}e^{-\lambda\phi}+\frac{5\rho_{m_{0}}\phi_{,a}}{2}+\rho_{m_{0}}a\phi_{,aa}-3a^{2}\phi_{,aa}+V_{0}e^{-\lambda\phi}a^{4}\phi_{,aa}\nonumber\\&+&\frac{\lambda V_{0}e^{-\lambda\phi}\phi_{,a}^{2}a^{4}}{2}=0\,\,,\nonumber\\
\end{eqnarray}
and for k=0,
\begin{eqnarray}  \label{26}
4V_{0}e^{-\lambda\phi}\phi_{,a}a^{3}+\frac{V_{0}e^{-\lambda\phi}\phi_{,a}^{3}a^{5}}{2}+\frac{\rho_{m_{0}}\phi_{,a}^{3}a^{2}}{4}+3\lambda a^{2}V_{0}e^{-\lambda\phi}+\frac{5\rho_{m_{0}}\phi_{,a}}{2}+\rho_{m_{0}}a\phi_{,aa}+\frac{\lambda V_{0}e^{-\lambda\phi}\phi_{,a}^{2}a^{4}}{2}+V_{0}e^{-\lambda\phi}a^{4}\phi_{,aa}=0\,\, ,\nonumber\\
\end{eqnarray}
\end{widetext}
where we consider $V(\phi)=V_{0}e^{-\lambda\phi}$ for the phantom like scalar field.

\begin{figure*}
\subfigure[Variation of $a$ with time]
{\includegraphics[width=81mm,height=52mm]{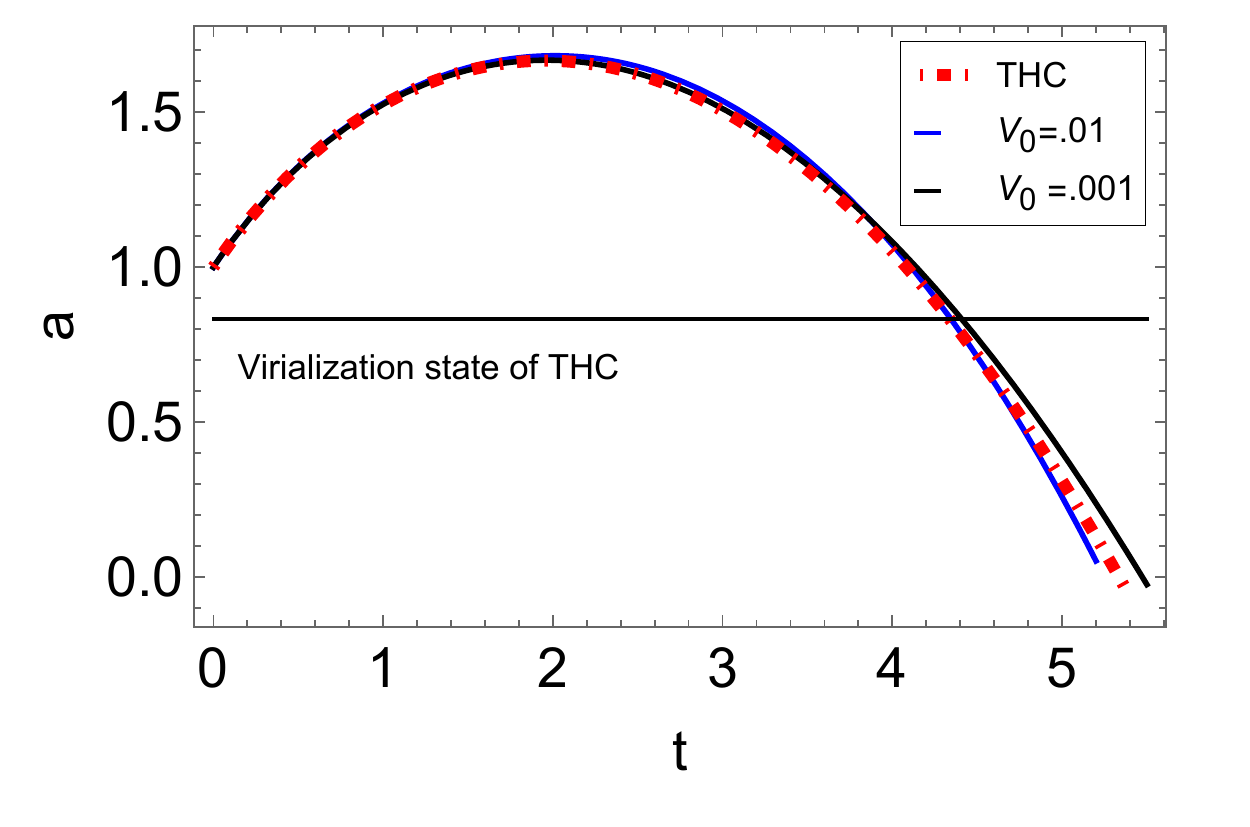}\label{Fiftha}}
\hspace{0.2cm}
\subfigure[Variation of $\omega_{\Phi}$ with time]
{\includegraphics[width=82mm,height=50mm]{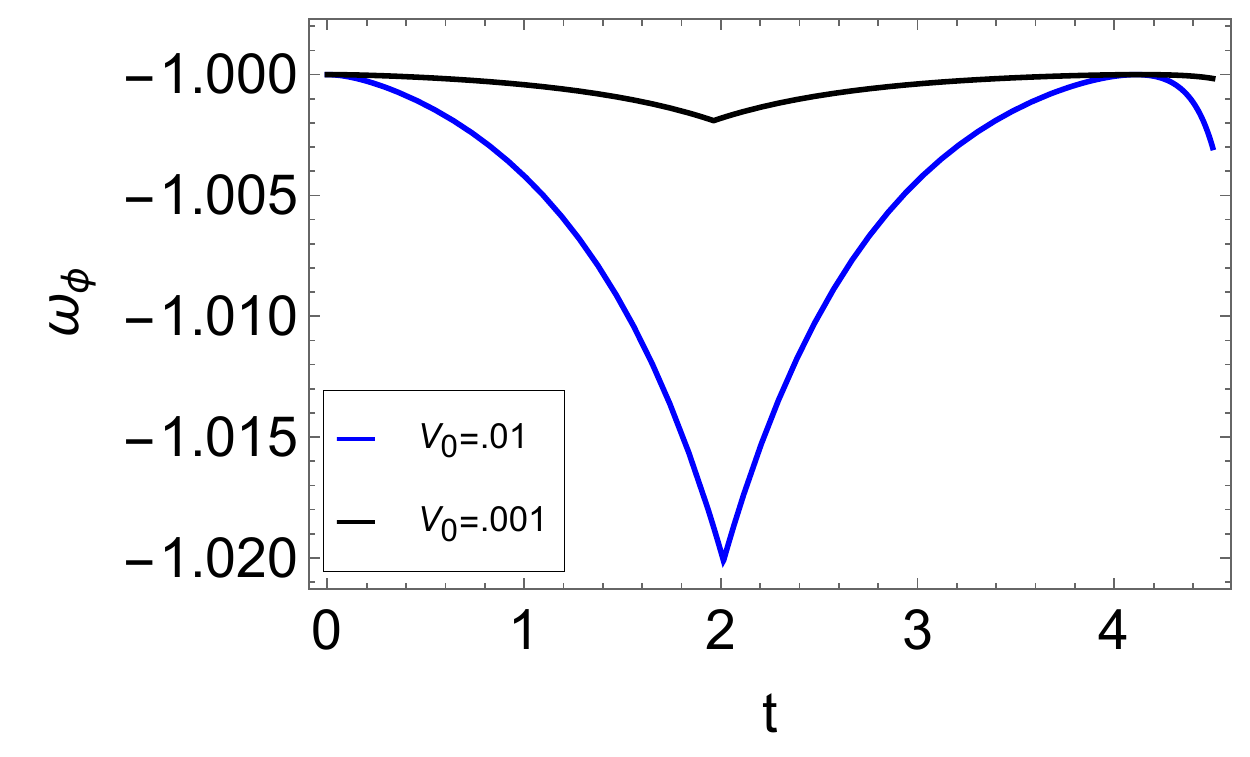}\label{Fifthb}}
\hspace{0.2cm}
\subfigure[Variation of $\frac{\rho_{\phi}}{\Bar{\rho}_{\phi}}$ with time]
{\includegraphics[width=82mm,height=50mm]{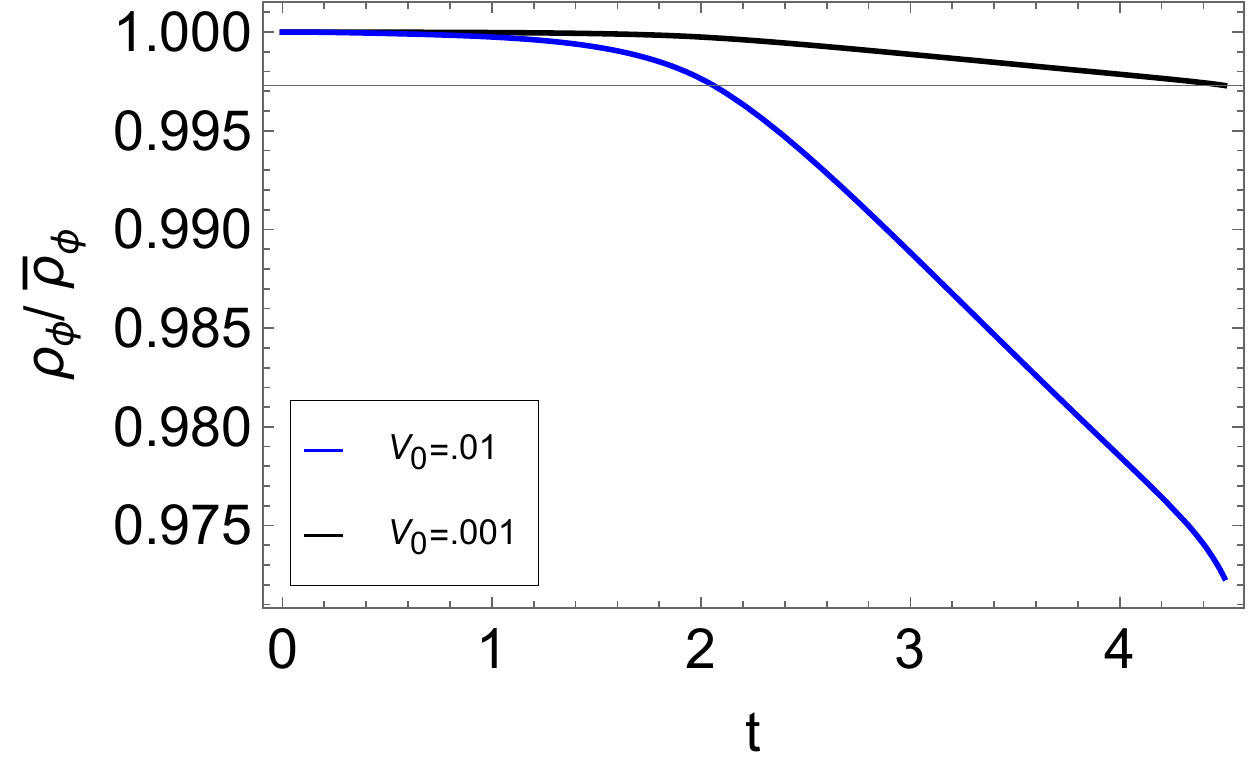}\label{Fifthc}}
\hspace{0.2cm}
\subfigure[Variation of $\omega_t$ with time]
{\includegraphics[width=85mm,height=54mm]{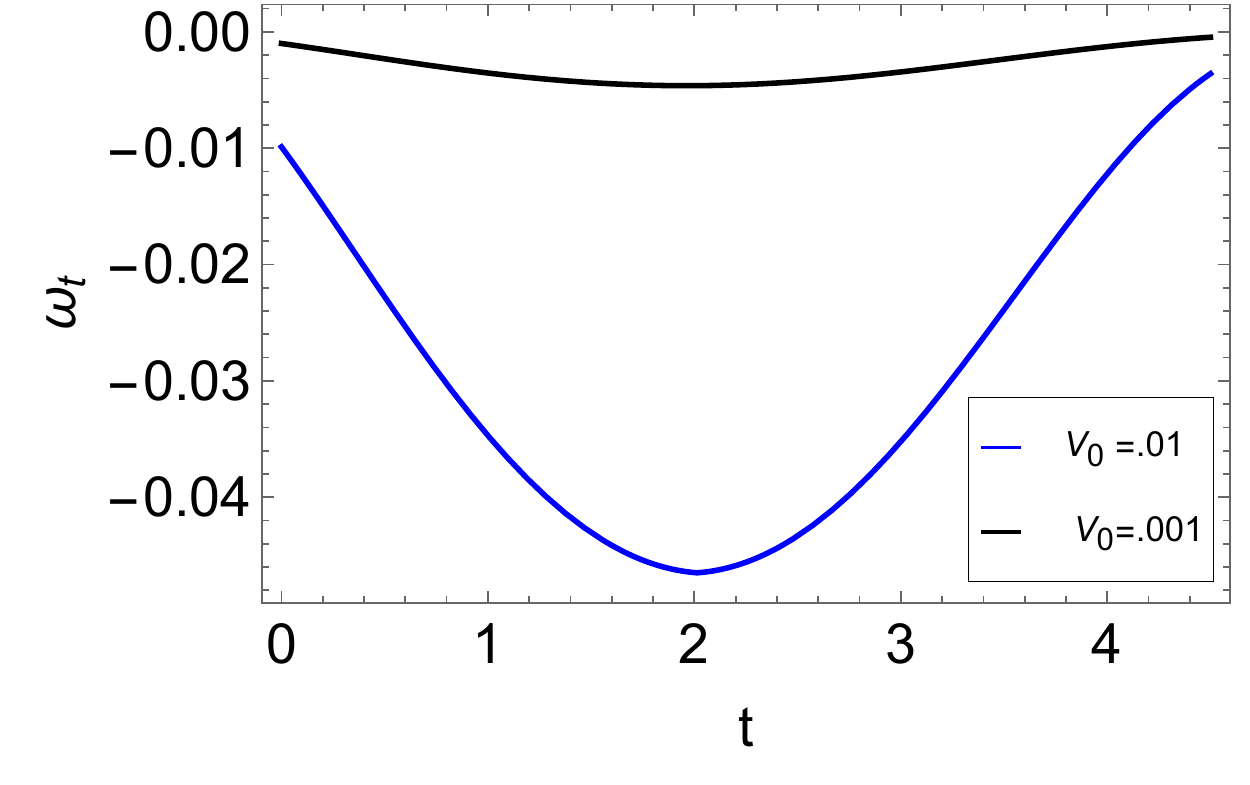}\label{Fifthd}}
\hspace{0.2cm}
\caption{Figure shows variation of different variables with variation of $V_{0}$ for scalar potential $V(\phi)=V_{0}e^{-\lambda\phi}$ for Phantom field.}
\label{Fifth}
\end{figure*}

\begin{figure*}
\subfigure[Variation of $a$ with time]
{\includegraphics[width=81mm,height=52mm]{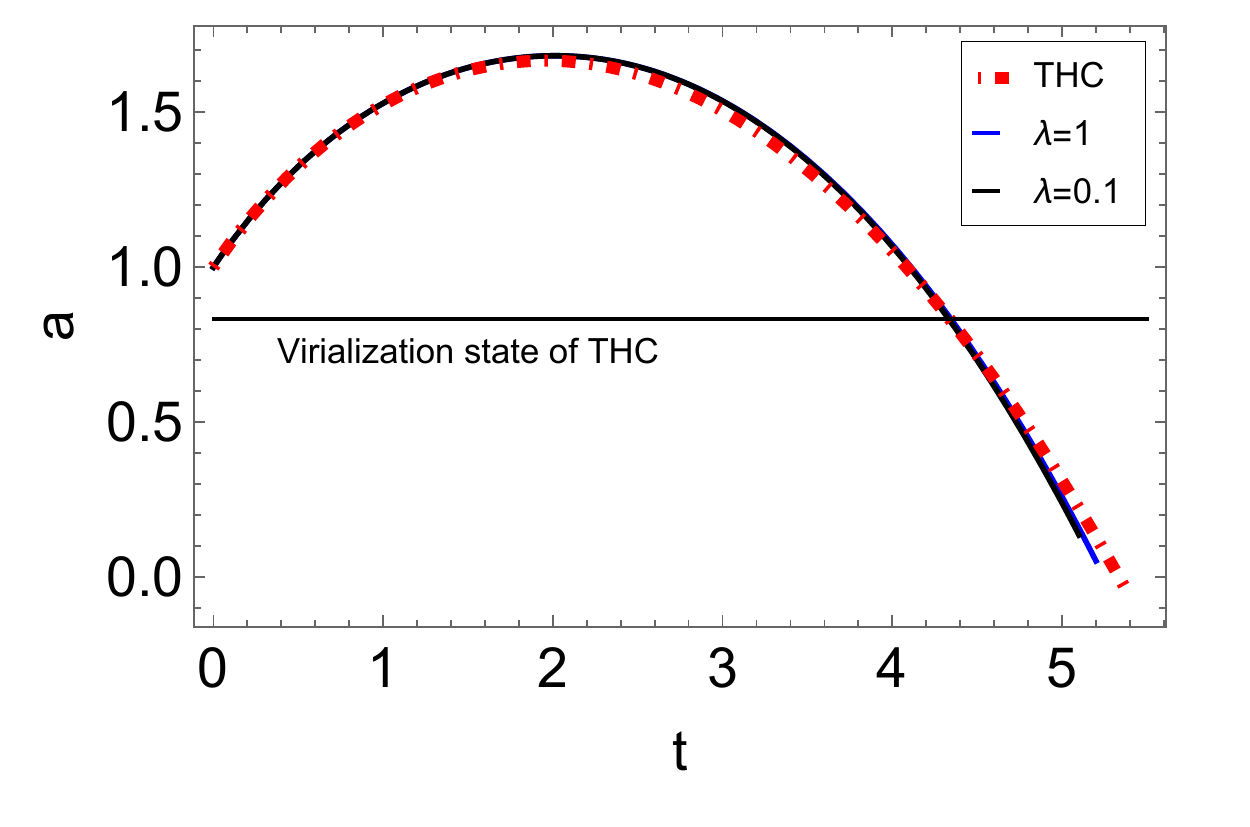}\label{Sixtha}}
\hspace{0.2cm}
\subfigure[Variation of $\omega_{\Phi}$ with time]
{\includegraphics[width=82mm,height=50mm]{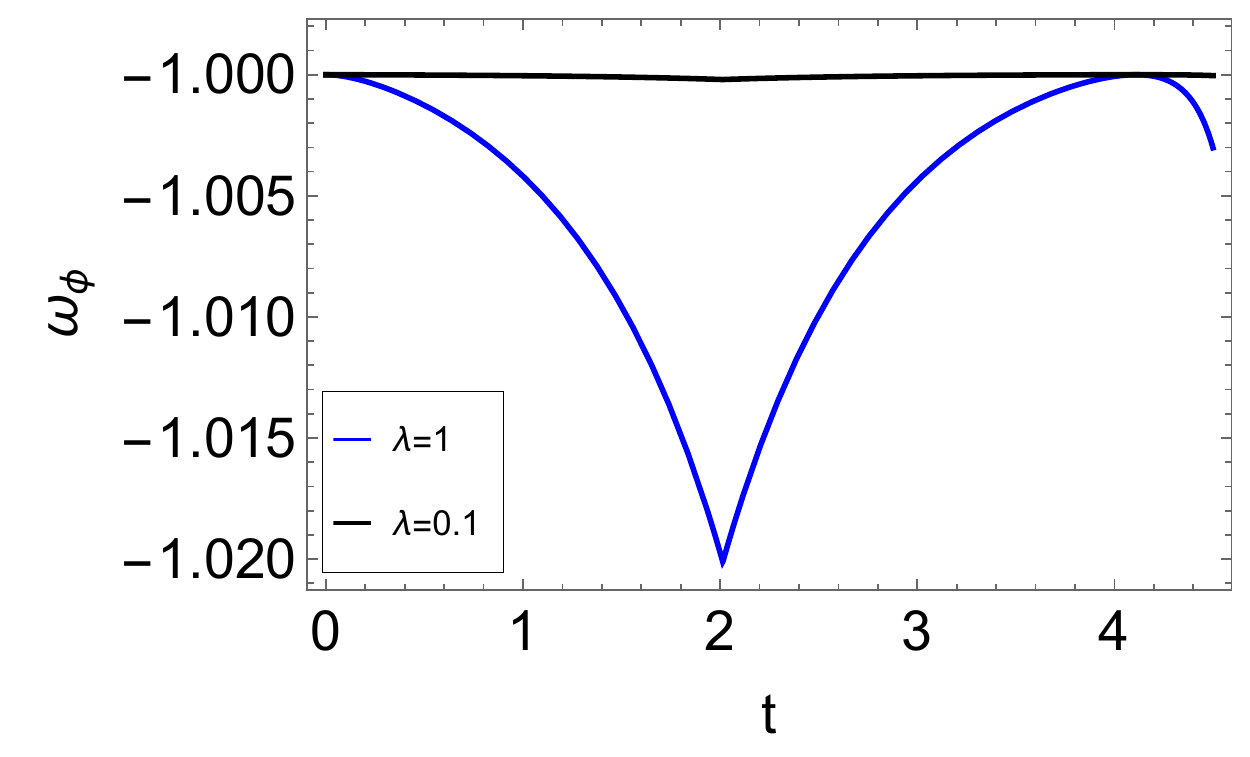}\label{Sixthb}}
\hspace{0.2cm}
\subfigure[Variation of $\frac{\rho_{\phi}}{\Bar{\rho}_{\phi}}$ with time]
{\includegraphics[width=82mm,height=50mm]{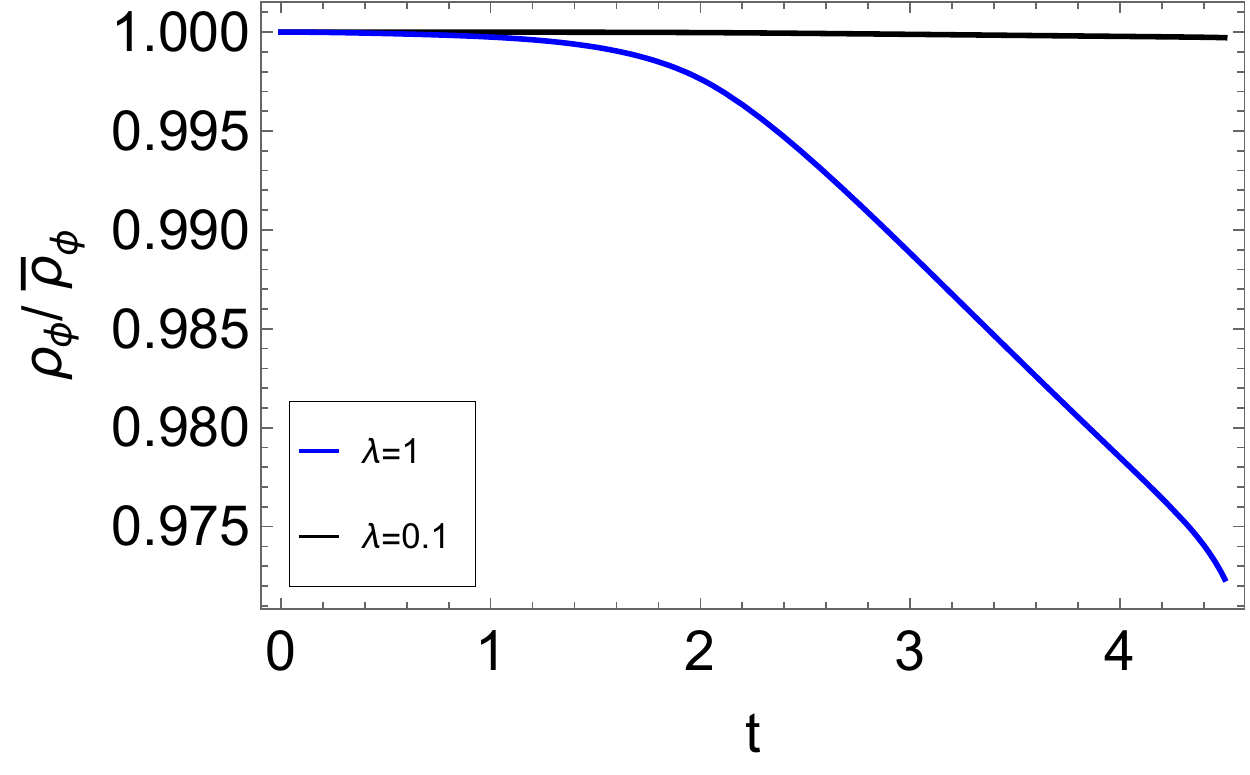}\label{Sixthc}}
\hspace{0.2cm}
\subfigure[Variation of $\omega_t$ with time]
{\includegraphics[width=85mm,height=54mm]{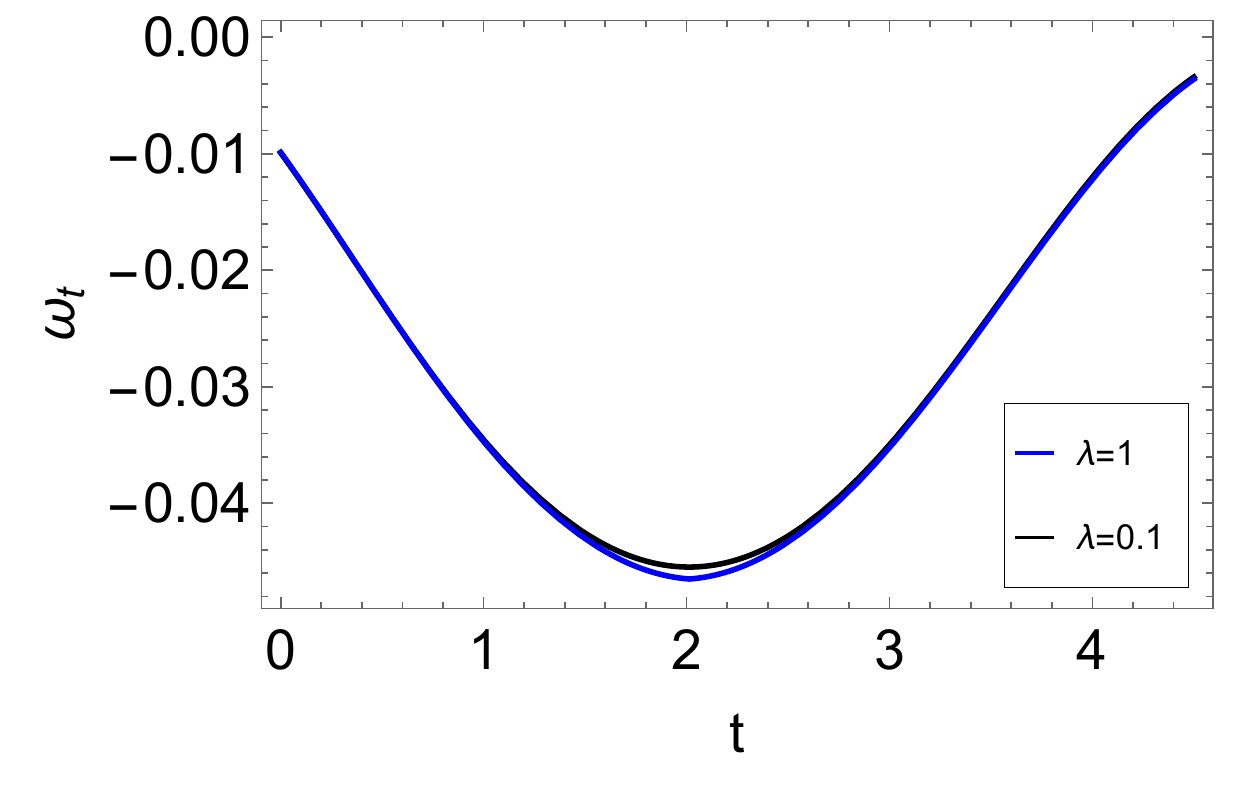}\label{Sixthd}}
\hspace{0.2cm}
\caption{Figure shows variation of different variables with variation of $\lambda$ for scalar potential $V(\phi)=V_{0}e^{-\lambda\phi}$ for Phantom field.}
\label{Sixth}
\end{figure*}

\begin{figure*}
\subfigure[Variation of $a$ with time for $\rho_{m_{0}}=5$]
{\includegraphics[width=81mm,height=52mm]{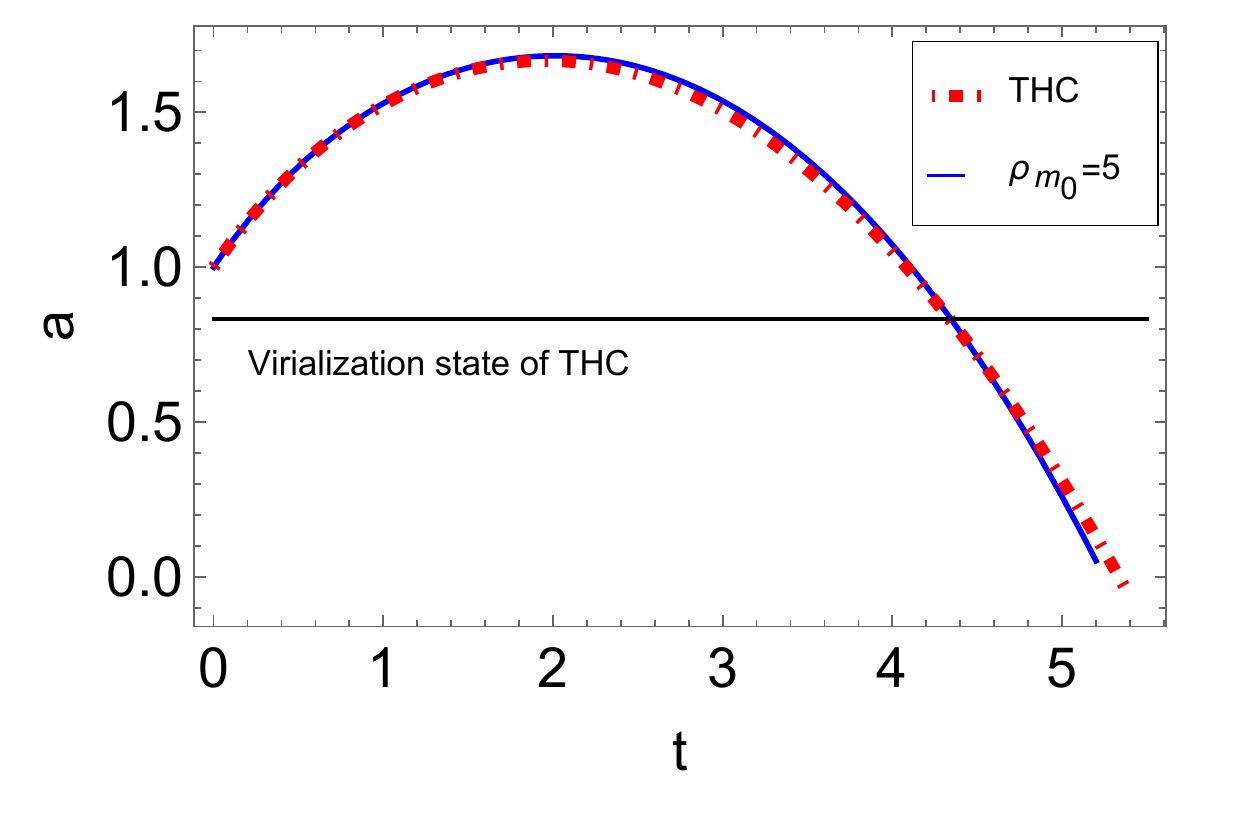}\label{Eightha}}
\hspace{0.2cm}
\subfigure[Variation of $a$ with time for $\rho_{m_{0}}=6$]
{\includegraphics[width=81mm,height=52mm]{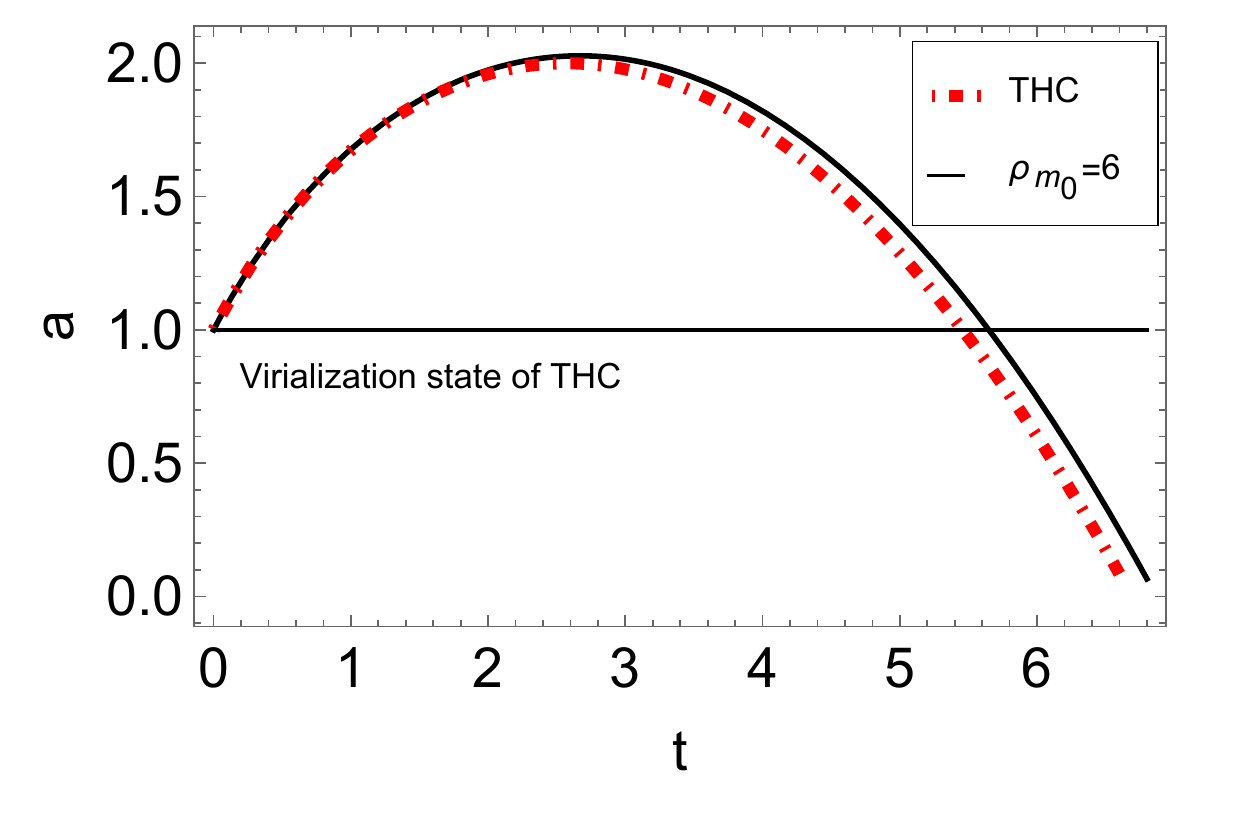}\label{Eighthb}}
\hspace{0.2cm}
\subfigure[Variation of $\omega_{\Phi}$ with time]
{\includegraphics[width=82mm,height=50mm]{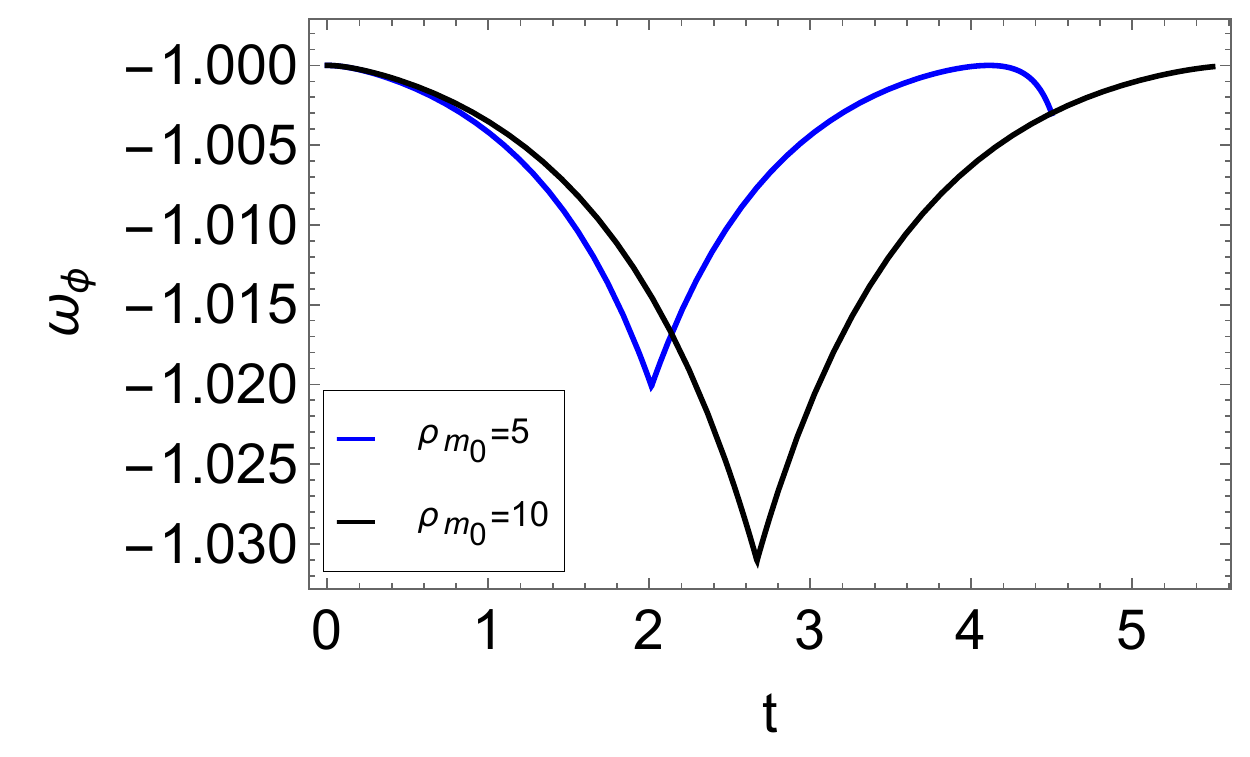}\label{Eighthc}}
\hspace{0.2cm}
\subfigure[Variation of $\frac{\rho_{\phi}}{\Bar{\rho}_{\phi}}$ with time]
{\includegraphics[width=82mm,height=50mm]{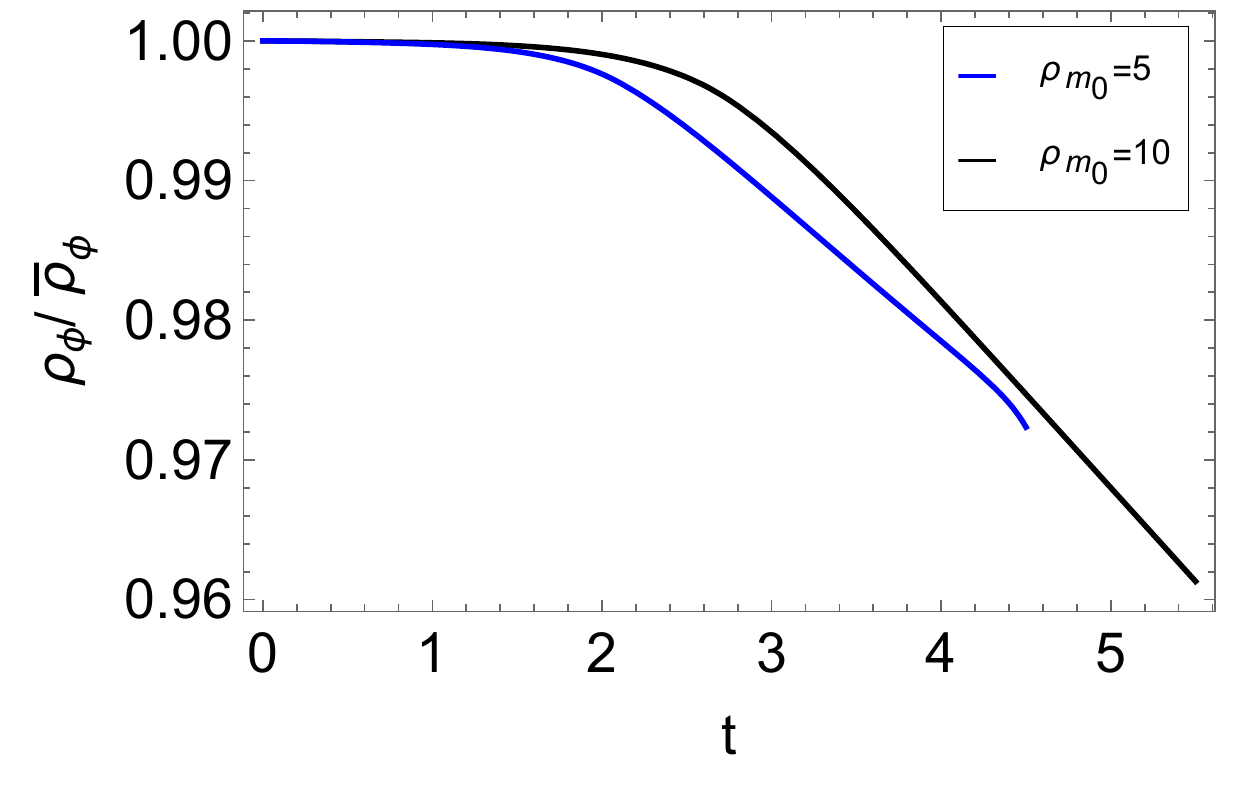}\label{Eighthd}}
\hspace{0.2cm}
\subfigure[Variation of $\omega_t$ with time]
{\includegraphics[width=85mm,height=54mm]{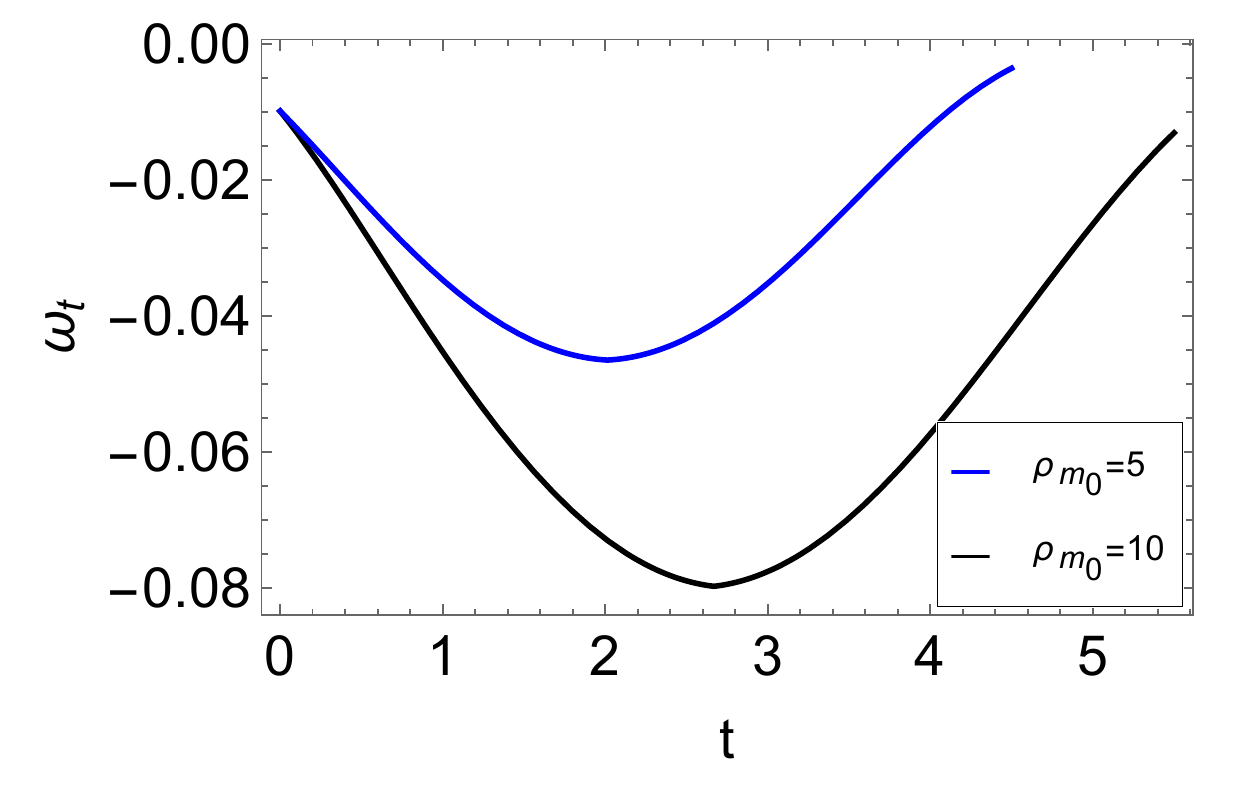}\label{Eighthe}}
\hspace{0.2cm}
\caption{Figure shows variation of different variables with variation of $\rho_{m_{_{0}}}$ for scalar potential $V(\phi)=V_{0}e^{-\lambda\phi}$ for Phantom field.}
\label{Eighth}
\end{figure*}
The differential Eqs.~(\ref{19}),(\ref{20}),(\ref{25}),(\ref{26}) are second order differential equations of $\phi(a)$. Therefore, we need to consider two initial conditions $\phi(a=1)$ and $\phi^{\prime}(a=1)$ to solve the differential equations.
Here we have taken the initial conditions as $\phi(a=1)=.001$, $\phi'(a=1)=0.00001$ for solving the differential Eqs.~(\ref{19}, \ref{25}).
We have three parameters $V_{0}$, $\rho_{m_{0}}$ and $\lambda$. 
In order to compare our model with the standard top-hat collapse model, in this paper, we only discuss those scenarios where the initial value of $\dot a$ is positive. The initial positive value of $\dot a$ ensures an initial expansion phase of the over-dense region. Now, depending on the values of the parameters $V_{0}$, $\rho_{m_{0}}$ and $\lambda$, the over-dense region may reach its maximum physical radius (i.e., at the turnaround time $t=t_{max}$) where from it starts collapsing. In Fig.~(\ref{regionplot1}) and Fig.~(\ref{regionplot2}), we show the parameters' space of $V_0$ and $\rho_{m_{0}}$ which allows the above-mentioned dynamics of the over-dense region in the presence of quintessence-like scalar field and phantom-like scalar field, respectively. In both cases, we consider $\lambda=1$.
The values of $V_0$ and $\rho_{m_{0}}$ in the unshaded region correspond to the ever-expanding dynamics of the over-dense patches.

Considering $k=1$ and the values of $V_0$ and $\rho_{m_0}$ from the shaded region of the Figs.~(\ref{regionplot1}, \ref{regionplot2}), it can be shown that the solution of the end state of the gravitational collapse of the two-fluid system is a space-time singularity. Therefore, in order to stabilize the system, like the standard top-hat collapse model, we invoke the Newtonian virialization technique in our model. In the top-hat collapse model, the matter in the over-dense sub-universe is pressureless, and therefore, as discussed before, it virializes when it reaches half of its maximum physical radius. However, as discussed before, when there exist two fluids inside a compact region, and if one of them is non-dust then the virialization radius may not be equal to half of the maximum physical radius. In Sec.~\ref{sec2}, we have briefly reviewed the works where the effect of dark energy on the virialization of dark matter is studied \cite{Lahav:1991wc, Steinh, Shapiro, Basilakos:2003bi, Caldwell:2003vq, Horellou:2005qc, Maor:2005hq, Percival:2005vm, Dey2, mota2006, Wang:2005ad, Basilakos:2006us, Maor:2006rh, Basilakos:2009mz, Basilakos:2010rs, Leewang, Chang:2017vhs}. In the next section, we show that the scalar field behaves almost like homogeneous dark energy in our model and therefore, we can use Eq.~(\ref{etaDE}) to calculate the vrialization radius of the over-dense region.

\section{Modeling of homogeneous dark energy scenario by the two-fluid model}
\label{sec5}

As we stated before, in our model the scalar field plays the role of dark energy, and the dust-like matter is considered dark matter. In this section, we show how the dynamics of the over-dense region vary when we change the values of $V_0, \rho_{m_{0}}$, and $\lambda$. Below we list the various dynamics of the over-dense region for different values of $V_0, \rho_{m_{0}}$, and $\lambda$.

In Fig.(\ref{First}), we show how the dynamical quantities like $a, \omega_{\phi}=p_\phi/\rho_\phi, \rho_\phi/\bar{\rho}_\phi$, and $\omega_t=p_\phi/(\rho_m + \rho_\phi)$ evolve with time for different values of $V_0$, where $\omega_{\phi}$ is the equation of state of the quintessence-like scalar field, $\rho_\phi/\bar{\rho}_\phi$ is the density-ratio between the energy densities of the scalar field in the over-dense region and background, and $\omega_t$ is the effective equation of state of the two-fluid system. In order to show, the dynamics of the above-mentioned dynamical quantities for different values of $V_0$, we consider  $\rho_{m_{0}}=5$, and $\lambda=1$. From Figs.~(\ref{regionplot1},\ref{regionplot2}), it can be understood that for a fixed value of $\rho_{m0}$ there exists a $\mathcal{V}_0$ such that for all values of $V_0<\mathcal{V}_0$, the over-dense region can have a collapsing phase after the initial phase of expansion. Therefore, for a fixed value of $\rho_{m_0}$, one cannot consider any arbitrarily large value of $V_0$ in order to model the desired top-hat collapse-like dynamics. The above statement is also true for $\rho_{m_0}$ since there exists a lower limit $\rho_{m_0}$ for a fixed value of $V_0$. Therefore, one cannot consider arbitrary large values of both the parameters $\rho_{m_0}$ and $V_0$ to model a top-hat collapse-like dynamics. Hence, in our model, we consider suitable small values of these two parameters. The plots of $a, \omega_{\phi}, \frac{\rho_\phi}{\bar{\rho}_\phi}$, and $\omega_t$ with respect to time for $V_0 = 0.01, 0.001$ are shown in Fig.~(\ref{Firsta}), (\ref{Firstb}), (\ref{Firstc}), (\ref{Firstd}), respectively. In Fig.~(\ref{Firstc}), it can be seen that the density ratio $\frac{\rho_\phi}{\bar{\rho}_\phi}$ slowly increases with comoving time and stays close to one throughout the total evolution of the over-dense region. The reason behind this increment of the value of the density ratio is that the background density of the scalar field decreases while the internal scalar field density approaches a constant value. However, one can consider suitable small values of $V_0$ to make the density ratio close to one throughout the evolution.      Therefore the quintessence-like scalar field in our model approximately behaves like homogeneous dark energy. On the other hand, from Fig.~(\ref{Firstb}), we can see that inside the over-dense region, the equation of state of the scalar field $\omega_{\phi}\sim -1$ throughout the evolution, and that is the reason why the internal density of the scalar field approaches to a constant value. Therefore, internally, the scalar field behaves like a cosmological constant $\Lambda$.  In Sec.~(\ref{sec2}), we discussed the homogeneous dark energy scenario where the dark energy is the cosmological constant. For this case, the solution of $\eta$
which is the ratio of virialized radius ($R_{Vir}$) and the turnaround radius ($R_{max}$) becomes:
\begin{eqnarray}
    \eta= 0.5 - 0.25 q -0.125 q^2+ \mathcal{O}(q^3)\,\,\nonumber ,
\end{eqnarray}
where $q = \left(\frac{\rho_{DE}}{\rho_{DM}}\right)_{a=a_{max}}$.
In our model, at the initial stage, we get $\rho_{\phi_{0}}=9.990\times10^{-4}$ and $\rho_{\phi_{0}}=9.990\times10^{-3}$ for $V_0=0.001$ and $V_0=0.01$, respectively. Therefore, initially, $\rho_{\phi_{0}}/\rho_{m_0}=1.998\times10^{-4}$ for $V_0=0.001$ and $\rho_{\phi_{0}}/\rho_{m_0}=1.998\times10^{-3}$ for $V_0=0.01$ which at the turnaround, becomes $4.09\times10^{-3}$ and $4.09\times10^{-2}$, respectively. Therefore, in our model, the value of $\eta$ does not differ much from that in the top-hat model where $\eta =0.5$. In the top-hat model, the time interval taken by the over-dense region to reach half of its maximum scale factor is 2.40. In our model, the time intervals are 2.52 and 2.79 for $V_0=0.001$ and $V_0=0.01$, respectively. Therefore, due to the effect of the quintessence-like scalar field, the over-dense region takes larger time to virialize. Fig.~(\ref{Firstd}) shows that the total or effective equation of state ($\omega_t$) of the two-fluid system stays close to zero throughout the evolution. It should be noted that here and throughout the remaining paper, we consider scale factor $a =1$ at the initial time to solve the differential Eqs.~(\ref{19}),(\ref{20}),(\ref{25}),(\ref{26}). 

In Fig.~(\ref{Third}), we show the evolution of $a, \omega_{\phi}, \frac{\rho_\phi}{\bar{\rho}_\phi}$, and $\omega_t$ for $\lambda =1$ and $\lambda = 0.1$. In this case, the values of $\rho_{m_0}$ and $V_0$ are fixed at $5$ and $0.001$, respectively. The Figs.~(\ref{Thirda}, \ref{Thirdb}, \ref{Thirdc}, \ref{Thirdd}) show similar type of behavior of $a, \omega_{\phi}, \frac{\rho_\phi}{\bar{\rho}_\phi}$, and $\omega_t$ as we have seen in the previous case. In this case, also the ratio $\frac{\rho_\phi}{\bar{\rho}_\phi}$ stays close to one, and the $\omega_{\phi} \sim -1$. Consequently, for different values of $\lambda$, we can still say our model approximately resembles the homogeneous dark-energy model. Here, for $\lambda =0.1$, at $t=0$, we get $\rho_{\phi_{0}}=9.999\times10^{-4}$. Therefore, at the initial stage, $\rho_{\phi_{0}}/\rho_{m_{0}}=1.999\times10^{-4}$ and at turnaround time, this ratio becomes $4.2\times10^{-3}$. Therefore, similar to the previous case, here also the value of $\eta\sim 0.5$ and the time interval taken by the over-dense region to reach the virialized radius is 2.69 that is $1.12$ times greater than the virialization time in top-hat collapse model.

The same similarity can be seen in Fig.~(\ref{Fourth}) where we vary the $\rho_{m_0}$. Therefore, observing the behavior of $a, \omega_{\phi}, \frac{\rho_\phi}{\bar{\rho}_\phi}$, and $\omega_t$ for all three cases, it can be concluded that our model of two-fluid system consisting of pressureless matter and quintessence-like scalar field approximately resembles the homogeneous dark energy model.

In Figs.~(\ref{Fifth}),(\ref{Sixth}), and (\ref{Eighth}), we show the dynamics of $a, \omega_{\phi}, \frac{\rho_\phi}{\bar{\rho}_\phi}$, and $\omega_t$ for various values of $V_0, \lambda$ and $\rho_{m_0}$ in the presence of phantom-like scalar field.
From Figs.~(\ref{Fiftha})-(\ref{Fifthd}), Figs.~(\ref{Sixtha})-(\ref{Sixthd}), and Figs.~(\ref{Eightha})-(\ref{Eighthd}), it can be seen that similar to the previous case here also $\frac{\rho_\phi}{\bar{\rho}_\phi}\sim 1$ and $\omega_{\phi} \sim -1$. Therefore, the phantom-like scalar field in our model also behaves like homogeneous dark energy. 

Till now what we have discussed deals with top-hat-like collapse in the presence of quintessence or phantom-like scalar fields. The ranges of potential parameter value and the initial matter density, which gives rise to such kind of collapse,  are shown in Figs.~(\ref{regionplot1},\ref{regionplot2}). What happens if the potential parameter value and the initial matter density does not lie in the shaded region of Figs.~(\ref{regionplot1},\ref{regionplot2})? In such cases we see that our model predicts that instead of gravitational collapse, the spherical over dense patch starts to expand. These patches expand forever producing void like structures, inside which the matter energy density is one order less than the background matter density for some period. Later the matter density goes down. For the phantom scalar fields, it is seen that the dark energy density grows inside the spherical expanding patch, when compared with the background dark energy density. For quintessence fields the dark energy density in the spherical patch becomes less than the corresponding energy density outside of the patch. If one assumes $\rho_{m_0}$ has a wide distribution in space for various non-linear perturbations then, for a fixed $V_0$ in the shaded regions of Figs.~(\ref{regionplot1},\ref{regionplot2}), one can have collapse or expansion depending on the value of  $\rho_{m_0}$. Our work predicts that some regions of the universe will collapse gravitationally whereas other regions will expand eternally to produce voids. For gravitational collapse of pressureless matter in the absence of scalar fields, one only obtains collapsing solutions.   
\section{Conclusion}
\label{sec6}

In this paper we have studied the gravitational dynamics of a two component system consisting of pressure-less matter and a scalar field, where the scalar field does not have any direct coupling with the matter component. The motivation to investigate this type of two-fluid dynamics is to understand how at a certain cosmological epoch, dark energy affects gravitational collapse of pressure-less dark matter, where the scalar field and the pressure-less matter play the role of dark energy and dark matter, respectively. We have chosen the scalar field potential in such a way that it represents the potential of quintessence or phantom like fields.  In order to model the dynamics of the primordial over-dense regions of dark matter in the presence of dark energy, we have chosen a closed FLRW metric as the internal space-time of the over-dense region which is seeded by the two component system. On the other hand, the background is modeled by flat FLRW metric which is also seeded by the two components: pressure-less matter and a scalar field. 

Previous authors have attempted this problem phenomenologically, where the guiding equation for the gravitational collapse of the dark matter component in presence of the scalar field was obtained from the Friedmann equations, but the complete relativistic framework was not used. The primary reason for not using the full general relativistic machinery is related to the fact that the dark energy component do not collapse with the dark matter part. In such a case one cannot use an isolated, closed FLRW spacetime which collapses towards a virial state. In the present work we have tried to implement a full general relativistic scheme to monitor the spherical collapse of the dark matter component in presence of the scalar field, up to virialization of the dark matter sector. To incorporate the relativistic treatment we have abandoned the idea of an isolated, closed FLRW spacetime collapse. Although we have used the FLRW spacetime with positive spatial curvature as the collapsing spacetime, we have matched this spacetime with an external, radiating Vaidyia spacetime at a suitable radial distance. In doing so the system has become an open system which can radiate.  In order to describe a matter flux through the boundary of the over-dense region, in the immediate neighborhood of the over-dense region, we consider an external generalized Vaidya space-time. We consider the potential of the scalar field $V(\phi)=V_0 e^{-\lambda\phi}$ which is the typical potential of quintessence-like and phantom-like scalar fields. We solve the Friedmann equations considering the above type of potential to investigate the dynamics of the over-dense region. In order to compare our results with that of the top-hat collapse model, we restrict ourselves to investigating those scenarios where the over-dense region collapses after an initial expansion phase. The collapsing spacetime is homogeneous and isotropic up to the matching radius, after which the spacetime remains isotropic but becomes inhomogeneous. 
 
In our scheme, the collapsing dark matter affects the dark energy sector locally and induces radiation in the Vaidya region. Our work predicts that there will be over-dense regions in the universe which will not collapse, they will expand forever producing voids. Which regions will collapse and which regions will not collapse depends upon the potential parameter $V_0$ and the initial value of the local dark matter density $\rho_{m_0}$. The nature of the outgoing flux, in the Vaidya region, will depend on whether there is a collapse or an expansion. Gravitational collapse in general always produce unclustered dark energy kind of a model, where  the dark energy density inside the collapsing core remains practically the same as that of the background dark energy density. On the other hand expanding patches can have clustering of dark energy as in these cases, the expanding patches have different dark energy density compared to the background spacetime.  The Vaidya radiation from collapsing regions are an unique prediction of our model and in near future we will like to work on the observational side of this problem.   
 
In this paper, we qualitatively discuss our model of spherical gravitational collapse of a two component system and do not attempt any comparison with observational data. One straightforward comparison can be done by comparing the theoretical value of the effective equation of state $w_t$ at the virialized state with the observed equation of state of the over-dense regions in the galactic cluster scale. This comparison would give constraints on the values of $V_0, \lambda$ and $\rho_{m_0}$ and that would be important to understand the effects of the homogeneous dark energy on structure formation at the galactic cluster scale. We will discuss this phenomenological aspect in the future. 

\section{Acknowledgement}

DD would like to acknowledge the support of
the Atlantic Association for Research in the Mathematical Sciences (AARMS) for funding the work.

\end{document}